\documentclass[12pt,a4paper,wide]{article}

\usepackage{graphicx}
\usepackage{amsfonts}
\usepackage{amsmath}
\usepackage{hyperref}
\usepackage{psfrag,subfigure,epsfig}

\usepackage[numbers,square,comma, compress]{natbib} 

\newcommand{\N}{\mathcal{N}}

\newcommand{\mb}{\mathbb}

\csname @addtoreset\endcsname{equation}{section}

\usepackage[lmargin=1.7 cm, rmargin=1.7 cm, tmargin=3.0cm, bmargin=3.0cm]{geometry}

\usepackage{float}

\title{\begin{flushright}{\vspace{-0.8cm}\small CERN-PH-TH/2013-223	\\[-12pt] \small MPP-2013-271}\end{flushright}
\vspace{0.8cm}
\bf{Non-Perturbative Nekrasov Partition Function from String Theory}}
\author{\Large  I.~Antoniadis\footnote{{\tt ignatios.antoniadis@cern.ch} \newline ${}^{}$~~~~
On leave from CPHT (UMR CNRS 7644) Ecole Polytechnique, F-91128 Palaiseau}~,~I. Florakis\footnote{\tt florakis@mppmu.mpg.de}~,~ 
S. Hohenegger\,\footnote{\tt stefan.hohenegger@cern.ch}~,\\[0.2cm] \Large K.S.~Narain\footnote{{\tt narain@ictp.trieste.it}}~ and A.~Zein Assi\footnote{\tt zeinassi@cern.ch}}
\date{}

\begin{document}

\maketitle

\begin{center}
\renewcommand{\thefootnote}{\fnsymbol{footnote}}\vspace{-0.5cm}
${}^{\footnotemark[1]\footnotemark[3]\footnotemark[5]}$ Department of Physics, CERN - Theory Division, CH-1211 Geneva 23, Switzerland\\[0.2cm]
${}^{\footnotemark[2]}$ Max-Planck-Institut f\"{u}r Physik,
	    Werner-Heisenberg-Institut,
              80805 M\"{u}nchen, Germany\\[0.2cm]
${}^{\footnotemark[4]}$ High Energy Section, The Abdus Salam International Center for Theoretical Physics,
\\Strada Costiera, 11-34014 Trieste, Italy\\[0.2cm]
${}^{\footnotemark[5]}$ Centre de Physique Th\'eorique (UMR CNRS 7644), Ecole Polytechnique, 91128 Palaiseau, France\\[0.2cm]
\end{center}

\abstract{
We calculate gauge instanton corrections to a class of higher derivative string effective couplings introduced in \cite{Antoniadis:2013bja}. We work in Type I string theory compactified on $K3\times T^2$ and realise gauge instantons in terms of D5-branes wrapping the internal space. In the field theory limit we reproduce the deformed ADHM action on a general $\Omega$-background from which one can compute the non-perturbative gauge theory partition function using localisation. This is a non-perturbative extension of \cite{Antoniadis:2013bja} and provides further evidence for our proposal of a string theory realisation of the $\Omega$-background.}

\newpage

\tableofcontents


\section{Introduction}
Gauge theory instantons have a natural description as bound states of D-branes in string theory \cite{Witten:1995gx,Douglas:1996uz}. This opens the possibility to calculate non-perturbative corrections in field theory from tree-level string amplitudes. Since the former are notoriously difficult to compute, whereas established techniques exist to obtain the latter, this provides very interesting insights into the inner structure of field theories and has lead to remarkable developments over the recent years. Furthermore, the series of works~\cite{Moore:1997dj,Lossev:1997bz,Nekrasov:2002qd,Nekrasov:2003rj} that led to the partition function of supersymmetric gauge theories in the $\Omega$-background has triggered considerable interest on this connection.

Indeed, four-dimensional super-Yang-Mills theory can be realised on a stack of D3-branes in Type IIB string theory, with additional D($-1$)-branes playing the role of instantons. Their corrections to the Yang-Mills action are captured by string disc diagrams with boundary field insertions. In~\cite{Billo:2006jm} the effect of a non-trivial constant string background in this setup was considered, by including additional bulk vertices in the tree-level amplitudes. It was shown that the insertion of anti-self-dual graviphoton field strength tensors in the point particle limit correctly reproduces the ADHM action on an $\Omega$-background with one of its deformation parameters switched off (\emph{e.g.} $\epsilon_+=0$). Using localisation techniques~\cite{Moore:1997dj,Lossev:1997bz,Nekrasov:2002qd,Nekrasov:2003rj,Losev:2003py,Iqbal:2003ix,Iqbal:2003zz} this allows one to compute the non-perturbative Nekrasov partition function $Z^{\textrm{Nek}}(\epsilon_{+}=0,\epsilon_{-})$.

Obtaining the partition function for a gauge theory on a generic $\Omega$-background (with $\epsilon_+\neq 0$) from string theory remains an interesting question. Phrased differently, one would like to find a  modification of the anti-self-dual graviphoton background, considered in \cite{Billo:2006jm}, giving rise to the fully deformed ADHM action in the point-particle limit of the appropriate disc diagrams. A hint for answering this question comes from a series of higher derivative one-loop couplings in the effective action of the Heterotic string compactified on $K3\times T^2$, considered in \cite{Antoniadis:2013bja}. These terms generalise a class of BPS-saturated couplings of gravitons and anti-self-dual graviphoton field strength tensors through additional couplings to the self-dual field strength tensor $F_{\bar{T}}$ of the vector partner of the K\"ahler modulus of the internal $T^2$. In the field theory limit, these one-loop amplitudes precisely reproduce the perturbative contribution of the Nekrasov 
partition function on the full $\Omega$-background, \emph{i.e.} with $\epsilon_{-},\epsilon_+\neq 0$. We therefore expect that, including $F^{\bar{T}}_{(+)}$ also as a background field in the instanton computation described above, allows us to extract the fully deformed ADHM action from string theory. In this paper, we show that this is indeed the case.

We work in Type I string theory compactified on $K3\times T^2$ and consider D9-branes together with D5-instantons wrapping $K3\times T^2$. This setting is dual to Heterotic string theory on $K3\times T^2$ and the corresponding background is given by anti-self-dual graviphotons and self-dual field strength tensors of the vector partner of $\bar{S}'$, which we refer to as $\bar{S}'$-vectors in the remainder of this work.  We compute all tree-level diagrams with boundary insertions in this background which, in the field theory limit, correctly reproduce the fully $\Omega$-deformed version of the ADHM action, which was used to compute Nekrasov's partition function~\cite{Nekrasov:2002qd,Nekrasov:2003rj}. 

This result can also be interpreted as computing gauge theory instanton corrections to the higher derivative couplings discussed in \cite{Antoniadis:2013bja}. The fact that we reproduce precisely the full Nekrasov partition function can also be seen as further evidence for the proposal that these couplings furnish a worldsheet description of the refined topological string. Indeed, our results suggest that the background introduced in \cite{Antoniadis:2013bja} can be understood as a physical realisation of the $\Omega$-background in string theory.

We would like to mention that various RR backgrounds were recently discussed in \cite{Ito:2010vx,Ito:2011cr} where the $\Omega$-deformed ADHM action was recovered using the language of D-instantons. However, contrary to our present work, the instanton calculation is performed without NS-NS field strengths. In addition, the interpretation in terms of a string effective coupling was lacking.

The paper is organised as follows. In Section \ref{Sect:InstantonsSetup}, we briefly review the construction of gauge theory instantons in Type I as a D5-D9 system and set the notation for the various moduli arising as massless open string states stretching between the various D-branes. In Section \ref{Sect:RefinedADHM}, we calculate the tree-level (disc) diagrams with bulk insertions of $\bar S'$-vectors. This is first done by introducing auxiliary fields as in \cite{Ito:2010vx,Ito:2011cr,Green:2000ke,Billo:2002hm,Billo:2006jm}. We show that, in the field theory limit, the resulting effective action precisely matches the one used in \cite{Nekrasov:2002qd} to derive the non-perturbative gauge theory partition function. However, correlation functions involving auxiliary fields are in general not well-defined since the latter are not BRST-closed. To this end, in Appendix \ref{App:PictureIndependence}, we recover the same effective couplings by inserting only physical states in the disc diagrams, hence justifying the result of the 
auxiliary field calculation. In Section~\ref{Sec:RefTop}, we discuss the connection of the present work with the higher derivative couplings studied in \cite{Antoniadis:2013bja}. Section~\ref{Sec:Conclusions} contains our conclusions.


\section{ADHM Instantons from String Theory}\label{Sect:InstantonsSetup}
We would like to use a string perspective to study gauge theory instantons. The latter are naturally realised as D-brane bound states and their contributions to the gauge theory action can be obtained at string tree-level. We work in Type I theory compactified on $\mathbb{R}^4\times T^2\times K3$. Gauge instantons in a theory living on a configuration\footnote{Before considering the effect of D5 instantons, one may start from a consistent string vacuum, \emph{e.g.} the model discussed in \cite{Bianchi:1990tb} with gauge group $U(16)\times U(16)$, where the total number of D5 branes (wrapping the space-time and $T^2$ directions) and D9 branes is fixed to be 16 by tadpole cancellation. The stack of 16 D9 branes may be further higgsed to $N<16$ by turning on appropriate Wilson lines along $T^2$, so that one may keep $N$ generic for the purposes of our analysis.} of $N$ D9-branes are the D5-instantons  wrapping  $K3 \times T^2$. For 
concreteness, we summarise the brane setup in the following table
\begin{center}
\parbox{14.5cm}{\begin{center}\begin{tabular}{|c||c||c|c|c|c||c|c||c|c|c|c|}\hline
&&&&&&&&&&&\\[-11pt]
\bf{brane} & \textbf{num.} & $X^0$ & $X^1$ & $X^2$ & $X^3$ & $X^4$ & $X^5$ & $X^6$ & $X^7$ & $X^8$ & $X^9$  \\\hline\hline
D9 & $N$ & $\bullet$ & $\bullet$ & $\bullet$ & $\bullet$ & $\bullet$ & $\bullet$ & $\bullet$ & $\bullet$ & $\bullet$ & $\bullet$\\\hline
D5 & $k$ &  &  &  &  & $\bullet$ & $\bullet$ & $\bullet$ & $\bullet$ & $\bullet$ & $\bullet$\\\hline
\end{tabular}\end{center}
${}$\\[-48pt]
\begin{align}
{}\hspace{2.9cm}\underbrace{\hspace{3.8cm}}_{\text{space-time}\,\sim\,\mathbb{R}^4}\underbrace{\hspace{1.9cm}}_{T^2}\underbrace{\hspace{3.75cm}}_{K3\,\sim\, T^4/\mathbb{Z}_2}\nonumber
\end{align}}
\end{center}
This configuration describes instantons\footnote{As usual, the back-reaction of the $D5$-instantons to the background is not considered.} of winding number $k$  in a gauge theory with $SU(N)$ gauge group. In order to see this, let us describe the massless spectrum of the theory. The notation used here is summarised in Appendix \ref{NoteConv}. More precisely, it is natural to decompose the 10-dimensional Lorentz group into 
\begin{align}
SO(10)\rightarrow SO(4)_{ST}\times SO(6)_{\text{int}}\rightarrow SO(4)_{\text{ST}}\times SO(2)_{T^2} \times SO(4)_{K3}\,,\label{DecomposeSo10}
\end{align}
reflecting the product structure of our geometry. In this way $I,J=1,\ldots,10$ denote indices of the full $SO(10)$, $\mu,\nu$ are indices of the space-time $SO(4)_{ST}$, with $\alpha,\beta$ ($\dot\alpha,\dot\beta$) the corresponding (anti-)chiral spinor indices, while $a,b$ denote internal $SO(6)_{\text{int}}$ indices.  For $SO(4)_{K3}\sim SU(2)_+\times SU(2)_-$, we introduce indices $A,B=1,2$ for fields transforming in the $(\mathbf{2},\mathbf{1})$ representation (positive chirality) and $\hat{A},\hat{B}=3,4$ for fields in the $(\mathbf{1},\mathbf{2})$ representation (negative chirality).  Following \cite{Billo:2002hm}, we associate upper indices of $SU(2)_+$ (resp. $SU(2)_-$) with charge $+1/2$ (resp. $-1/2$) of $SO(2)_{T^2}$ and downstairs indices with charge $-1/2$ (resp. $+1/2$) respectively. In this way, internal indices cannot be raised and lowered with the help of $\epsilon$-tensors, but we have to keep track of their position.

In this setting, there are three kinds of open string sectors which are relevant for our subsequent discussion. They can be characterised according to the location of their endpoints.

\begin{enumerate}
\item 9-9 sector\\
The massless excitations consist of a number of $\mathcal{N}=2$ vector multiplets, each of which containing a vector field $A^\mu$, a complex scalar $\phi$ as well as four gaugini $(\Lambda^{\alpha A}, \Lambda_{\dot\alpha A})$ in the $(\mathbf{2},\mathbf{1})$ representation of $SU(2)_{+}\times SU(2)_-$ with $SO(2)_{T^2}$ charges $(1/2,-1/2)$ respectively. The bosonic degrees of freedom stem from the NS sector, while the fermonic ones from the R sector. These fields separately realise a Yang-Mills theory living on the four-dimensional space-time. 
\item 5-5 sector\\
These states are moduli (i.e. non dynamical fields) from a string perspective, due to the instantonic nature of the corresponding D5-branes. Indeed, the states in this sector cannot carry any space-time momentum because of Dirichlet boundary conditions in all directions except along $K3\times T^2$. From the Neveu-Schwarz (NS) sector, we have six bosonic moduli, which we  write as a real vector $a^\mu$ and a complex scalar $\chi$. From the point of view of the SYM theory living on the world-volume of the D9-branes, $a^\mu$ corresponds to the position of gauge theory instantons. From the Ramond sector, we have eight fermionic moduli, which we denote as $M^{\alpha A}$, $\lambda_{\dot\alpha A}$.
\item 5-9 and 9-5 sectors\\
Also this sector contains moduli from a string point of view. From the NS sector, the fermionic coordinates have integer modes giving rise to two Weyl spinors of $SO(4)_{ST}$ which we call $(\omega_{\dot\alpha},\bar\omega_{\dot\alpha})$. Notice that these fields all have the same chirality, which in our case is anti-chiral, owing to a specific choice of reflection rules in the orbifold construction (see \cite{Billo:2002hm} for a similar discussion in a related setup). From a SYM point of view, these fields control the size of the instanton. In the R sector, fields are half-integer moded giving rise to two fermions $(\mu^A, \bar\mu^A)$ transforming in the $(\mathbf{2},\mathbf{1})$ representation of $SU(2)_+$ with positive charge under $SO(2)_{T^2}$.
\end{enumerate}
For the reader's convenience, the field content is compiled in Table~\ref{Tab:Fields}.
\begin{table}[htbp]
\begin{center}
\begin{tabular}{|c||c||c|c|c|c|c|}\hline
&&&&&&\\[-11pt]
\bf{sector} & \bf{field} & $SO(4)_{ST}$ & $c_{T^2}$ & $SU(2)_+\times SU(2)_-$ & \bf{statistic} & \bf{R / NS} \\\hline\hline
&&&&&&\\[-11pt]
9-9 & $A^\mu$ & $(\mathbf{1/2},\mathbf{1/2})$ & $0$ & $(\mathbf{1},\mathbf{1})$ & boson & NS \\
&&&&&&\\[-11pt]\hline
&&&&&&\\[-11pt]
 & $\Lambda^{\alpha A}$ & $(\mathbf{1/2},\mathbf{0})$ & $1/2$ & $(\mathbf{2},\mathbf{1})$ & fermion & R \\
 &&&&&&\\[-11pt]\hline
 &&&&&&\\[-11pt]
 & $\Lambda_{\dot{\alpha} A}$ & $(\mathbf{0},\mathbf{1/2})$ & $-1/2$ & $(\mathbf{2},\mathbf{1})$ & fermion & R \\
 &&&&&&\\[-11pt]\hline
 &&&&&&\\[-11pt]
 & $\phi$ & $(\mathbf{0},\mathbf{0})$ & $-1$ & $(\mathbf{1},\mathbf{1})$ & boson & NS \\ &&&&&&\\[-11pt]\hline\hline
 &&&&&&\\[-11pt]
 5-5 & $a^\mu$ & $(\mathbf{1/2},\mathbf{1/2})$ & $0$ & $(\mathbf{1},\mathbf{1})$ & boson & NS \\
&&&&&&\\[-11pt]\hline
&&&&&&\\[-11pt]
 & $\chi$ & $(\mathbf{0},\mathbf{0})$ & $-1$ & $(\mathbf{1},\mathbf{1})$ & boson & NS \\
&&&&&&\\[-11pt]\hline
&&&&&&\\[-11pt]
& $M^{\alpha A}$ & $(\mathbf{1/2},\mathbf{0})$ & $1/2$ & $(\mathbf{2},\mathbf{1})$ & fermion & R \\
&&&&&&\\[-11pt]\hline
&&&&&&\\[-11pt]
& $\lambda_{\dot{\alpha} A}$ & $(\mathbf{0},\mathbf{1/2})$ & $-1/2$ & $(\mathbf{2},\mathbf{1})$ & fermion & R \\
&&&&&&\\[-11pt]\hline\hline
&&&&&&\\[-11pt]
5-9 & $\omega_{\dot{\alpha} }$ & $(\mathbf{0},\mathbf{1/2})$ & $0$ & $(\mathbf{1},\mathbf{1})$ & boson & NS \\
&&&&&&\\[-11pt]\hline
&&&&&&\\[-11pt]
 & $\mu^A$ & $(\mathbf{0},\mathbf{0})$ & $1/2$ & $(\mathbf{2},\mathbf{1})$ & fermion & R \\[2pt]\hline
\end{tabular}
\end{center}
\caption{Overview of the massless open string spectrum relevant for the disc amplitude computation. We display the transformation properties under the groups $SO(4)_{ST}\times SU(2)_+\times SU(2)_-$, while $c_{T^2}$ is the charge under $SO(2)_{T^2}$. The last two columns denote whether the field is bosonic or fermonic and the sector it stems from.}
\label{Tab:Fields}
\end{table}
As was discussed in \cite{Green:2000ke,Billo:2002hm}, the string tree-level effective action involving only the massless vector multiplets of the 9-9 sector exactly reproduces in the field theory limit pure $\mathcal{N}=2$ super-Yang-Mills theory with $SU(N)$ gauge group. Inclusion of the remaining moduli fields gives rise to the ADHM action describing instantonic corrections with instanton\footnote{When taking the field theory limit, one should pay attention to the dimensionality of the various fields. In particular, a rescaling of the ADHM moduli is necessary in order for the field theory limit to be well-defined as an appropriate double scaling limit in which $g_{\textrm{YM}}$ is held fixed.} number $k$. Therefore, this setup provides a stringy description of the gauge theory instantons. 

Furthermore, by coupling the theory to a constant anti-self-dual graviphoton background \cite{Billo:2006jm}, the resulting effective action coincides with the ADHM action in the $\Omega$-background used in \cite{Nekrasov:2002qd} in the case where one of the deformation parameters vanishes (say, $\epsilon_+=0$). While the ADHM action is exact under a nilpotent Q-symmetry, the latter is still present after the deformation with $\epsilon_+$. Hence, one can use localization techniques in order to compute the instanton partition function \cite{Nekrasov:2002qd}. 

From a practical perspective, this deformation is obtained by computing string disc diagrams with bulk insertions of the anti-self-dual graviphoton. Due to its anti-self-duality, the only instanton contributions come from the diagrams with insertions from the 5-5 sector and no mixed diagrams\footnote{By mixed diagrams we refer to disc diagrams whose boundary lies on both D9- and D5-branes.} contribute.
\begin{figure}[h!t]
\begin{center}
\parbox{8cm}{\epsfig{file=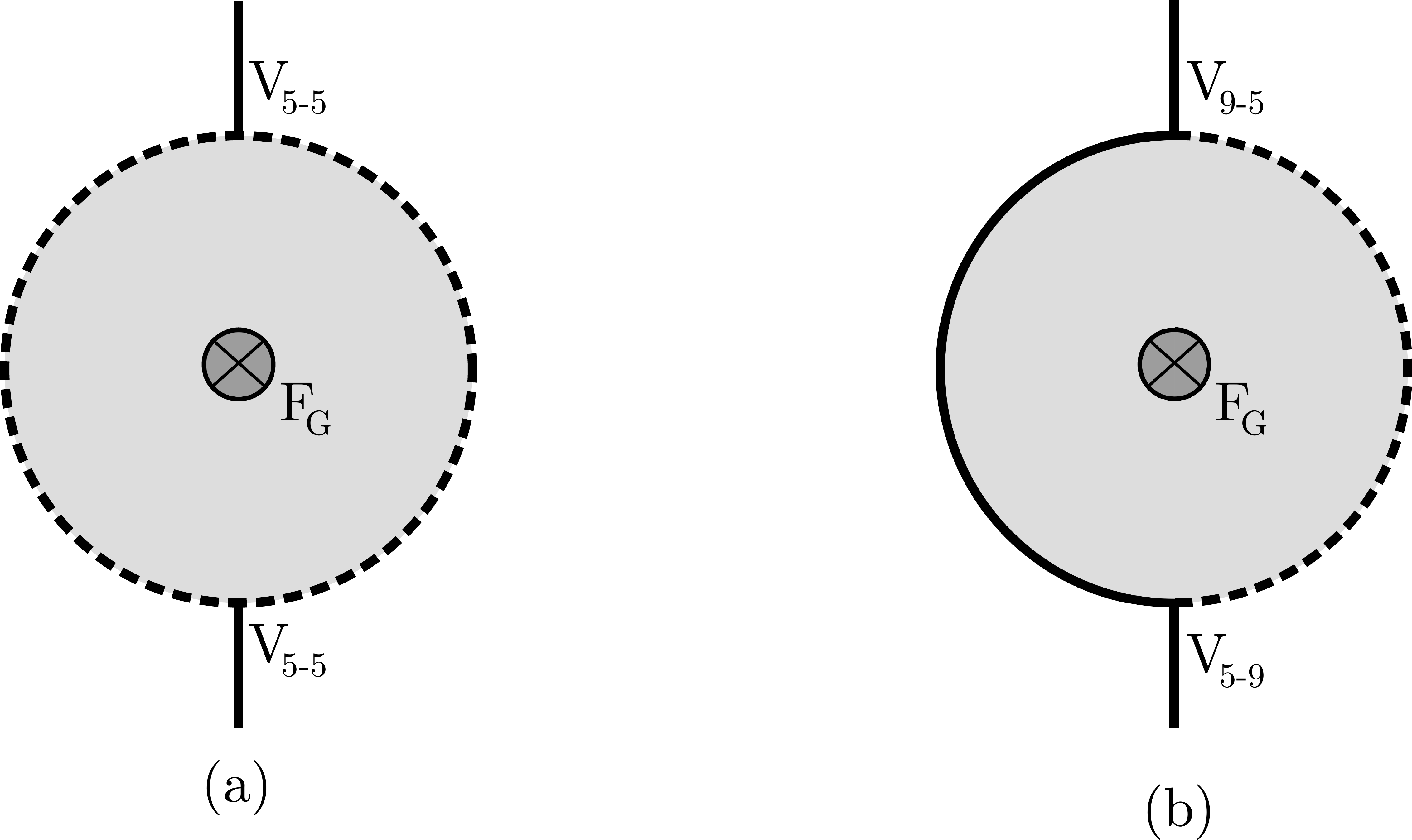,width=7.5cm}}
\end{center}
\hspace{1cm}\parbox{0.89\textwidth}{\caption{\small{\it Three-point disc diagrams with graviphoton bulk-insertion. Diagram (a) involves two boundary insertions from the 5-5 sector, whereas diagram (b) two insertions from the 5-9 sector. While the whole boundary of diagram (a) lies on the D5-branes, diagram (b) lies partly on the D9- and partly on the D5-branes. Notice that the latter mixed boundary conditions appear only in the space-time directions.}}}
\label{fig-ee}
\end{figure}

In the following section, we generalise this construction to the case where the background includes, in addition to the anti-self-dual graviphoton field strength, the self-dual field strength of the $\bar{S}'$-vector. Such a background was recently introduced in \cite{Antoniadis:2013bja} as a proposal for a worldsheet description of the refined topological string. We show that this generalised background reproduces the ADHM action in the presence of a general $\Omega$-background, therefore providing a non-perturbative check of the proposal of \cite{Antoniadis:2013bja}.


\section{Refined ADHM Instantons in String Theory}\label{Sect:RefinedADHM}
In this section we calculate disc diagrams relevant for making contact with the instanton sector of gauge theory using the RNS formalism. We use a method similar to the one employed in \cite{Billo:2006jm}, namely, we introduce auxiliary fields which linearise the superalgebra. As a consequence all diagrams involve only a single bulk insertion of the  $\bar{S}'$ field strength tensor and two insertions of boundary fields. Even though the computation is considerably simplified in this case, one faces the problem that auxiliary fields are not physical and, therefore, not BRST-closed. String theory amplitudes involving such fields are generically not independent of the superghost picture, thus rendering the results a priori ambiguous (an example of this phenomenon is given in Section~\ref{Sect:PictureDependence}). Nevertheless, in our case, the correct result is obtained for the particular choice of picture, which we use in this section. Indeed, in Appendix~\ref{App:PictureIndependence} we show that one may 
repeat the calculation in a physical framework (i.e. without resorting to auxiliary fields), where picture independence is restored. The justification for our simplified computation is provided in the same appendix. There we show that, for this special choice of picture, the relevant disc diagrams factorise in a particular form which can be re-interpreted in terms of (unphysical) auxiliary fields and matches precisely the result obtained using auxiliary fields.

\subsection{Vertex Operators and Disc Diagrams}\label{Sect:DiskDiagrams}

We denote the ten-dimensional bosonic and fermonic worldsheet fields collectively as $X^I$ and $\psi^I$ respectively, with $I=1,\ldots,10$. More precisely, we use $(X^{\mu},\psi^{\mu})$, $(Z,\Psi)$ and $(Y^i,\chi^i)$ for worldsheet fields along the four-dimensional space-time, $T^2$ and $K3$ directions respectively. In the following, for simplicity, we consider an orbifold representation of  $K3$ as $T^4/\mathbb{Z}_2$. However, we expect our results to be valid also for a generic (compact) $K3$. The vertex operators relevant for the disc amplitudes involve the ADHM moduli appearing in the massless spectrum. From the 5-5 sector, we need
\begin{align}
 V_{a}(z)&= g_6\,a_{\mu}\psi^{\mu}(z)e^{-\varphi(z)}\,,\\
 V_{\chi}(z)&= \frac{\chi}{\sqrt{2}}\bar\Psi(z)e^{-\varphi(z)}\,,\\
 V_{M}(z)&= \frac{g_6}{\sqrt{2}}\,M^{\alpha A}S_{\alpha}(z)S_{A}(z)e^{-\tfrac{1}{2}\varphi(z)}\,.
\end{align}
From the 5-9 and 9-5 sectors, we use
\begin{align}
 V_{\omega}(z)=& \frac{g_6}{\sqrt{2}}\,\omega_{\dot\alpha}\Delta(z)S^{\dot\alpha}(z)e^{-\varphi(z)}\,,\\
 V_{\bar\omega}(z)=& \frac{g_6}{\sqrt{2}}\,\bar\omega_{\dot\alpha}\bar\Delta(z)S^{\dot\alpha}(z)e^{-\varphi(z)}\,.
\end{align}
Here $\Delta(z),\,\bar\Delta(z)$ are twist and anti-twist fields with conformal weight $1/4$, which act by changing the boundary conditions and $g_6$ is the D5-instanton coupling constant. The superghost is bosonised in terms of the free boson $\varphi$. In order to linearize the supersymmetry transformations, a set of auxiliary fields is introduced whose vertex operators read as follows:
\begin{align}
 V_{Y}(z)&=\sqrt{2}\,g_6\,Y_{\mu}\bar\Psi(z)\psi^{\mu}(z)\,,&&\,V_{Y^{\dag}}(z)=\sqrt{2}\,g_6\,Y_{\mu}^{\dag}\Psi(z)\psi^{\mu}(z)\,,\label{Auxiliary1}\\
 V_{X}(z)&=g_6\, X_{\dot\alpha}\Delta(z)S^{\dot\alpha}\bar\Psi(z)\,,&&\,V_{X^{\dag}}(z)=g_6\, X_{\dot\alpha}^{\dag}\Delta(z)S^{\dot\alpha}\Psi(z)\,,\\
 V_{\bar X}(z)&=g_6\, \bar X_{\dot\alpha}\bar\Delta(z)S^{\dot\alpha}\bar\Psi(z)\,,&&\,V_{\bar X^{\dag}}(z)=g_6\, \bar X_{\dot\alpha}^{\dag}\bar\Delta(z)S^{\dot\alpha}\Psi(z)\,.\label{Auxiliary3}
\end{align}
Finally, we turn to the closed string background defined in \cite{Antoniadis:2013bja}. The vertex operator of the anti-self-dual graviphoton in the $(-1)$-ghost picture at zero momentum is given by:
\begin{align}
	V^{F^G}(y,\bar y)=\frac{1}{8\pi\,\sqrt{2}}F_{\mu\nu}^{G} \bigg[\,\psi^{\mu}\psi^{\nu}(y)e^{-\varphi(\bar y)}{\bar\Psi}(\bar y)\,+e^{-\varphi(y)}\bar\Psi(y)\psi^{\mu}\psi^{\nu}(\bar y)&\nonumber\\
-\frac{i}{2}\, e^{-\tfrac{1}{2}(\varphi(y)+\varphi(\bar y))} S_{\alpha}(y) (\sigma^{\mu\nu})^{\alpha\beta} 
S_{\beta}(\bar y)\,\epsilon^{AB}\,S_{A}(y)S_{B}(\bar y)\bigg]&\,.\label{GraviphotonVertex}
\end{align}
The vertex operator for the self-dual field strength tensor of the $\bar{S}'$-vector in the $(-1)$-ghost picture and at zero momentum is given as a sum of an NS-NS part (first line) and a R-R part (second line) \cite{Antoniadis:2013bja}:
\begin{align}\label{UbarVec}
V^{F^{\bar{S}'}}(y,\bar y)=\frac{1}{8\pi\,\sqrt{2}}F_{\mu\nu}^{\bar{S}'} \bigg[\,\psi^{\mu}\psi^{\nu}(y)e^{-\varphi(\bar y)}{\bar\Psi}(\bar y)\,+e^{-\varphi(y)}\bar\Psi(y)\psi^{\mu}\psi^{\nu}(\bar y)&\nonumber\\
+\frac{i}{2}\, e^{-\tfrac{1}{2}(\varphi(y)+\varphi(\bar y))} S_{\dot{\alpha}}(y) (\bar{\sigma}^{\mu\nu})^{\dot{\alpha}\dot{\beta}} 
S_{\dot{\beta}}(\bar y)\,\epsilon_{\hat A\hat B}\,S^{\hat A}(y)S^{\hat B}(\bar y)\bigg]&\,.
\end{align}
The disc diagrams involving the bulk insertions of the RR part of the anti-self-dual graviphoton \eqref{GraviphotonVertex} have already been extensively studied in the literature (see, for example \cite{Billo:2006jm}). Following the analysis performed below, one can show that the NS-NS part also gives the same contribution leading to the $\epsilon_{-}$-dependent part of the $\Omega$-deformed ADHM action. Here, we focus on the more interesting case of the self-dual insertions of $\bar{S}'$-field strengths. The relevant disc diagrams correspond to the following correlators:
\begin{align}
 D_{Y^{\dagger}a F^{\bar S'}}(z_1,z_2,y,\bar{y})&=\left<V_{Y^{\dag}}(z_1)\,V_a(z_2)\,V_{F^{\bar{S}'}}(y,\bar y)\right>_{\text{Disc}}\,,\label{UnmixedDiag}\\
 D_{\bar{X}^\dagger\omega F^{\bar S'}}(z_1,z_2,y,\bar{y}) &=\left<V_{\bar X^{\dag}}(z_1)\,V_{\omega}(z_2)\,V_{F^{\bar{S}'}}(y,\bar y)\right>_{\text{Disc}}\,,\label{MixedDiag}\\
 D_{MM F^{\bar S'}}(z_1,z_2,y,\bar{y})&=\left<V_{M}(z_1)\,V_{M}(z_2)\,V_{F^{\bar{S}'}}(y,\bar y)\right>_{\text{Disc}}\,.\label{UnmixedFerm}
\end{align}


\subsubsection*{$D_5-D_5$ disc diagrams}

We start by evaluating the amplitude \eqref{UnmixedDiag} corresponding to a disc diagram whose boundary lies entirely on the stack of D5-branes and with the $\bar{S}'$-modulus inserted in its bulk. This allows us to fix the normalisation of the $\bar S'$ vertex operator, which is needed to match precisely all numerical coefficients in the refined ADHM action. The amplitude we consider is given by
\begin{align}
 \left<\left<V_{Y^{\dag}}\,V_a\,V_{F^{\bar S'}}\right>\right>_{\textrm{Disc}}= \frac{8\pi^2}{g_{\textrm{YM}}^2}\int\frac{dz_1\, dz_2\, dy \,d\bar y}{dV_{\textrm{CKG}}}D_{Y^{\dagger}a F^{\bar S'}}(z_1,z_2,y,\bar{y})\,.\label{Corr11}
\end{align}
Here the double brackets define the integral of the correlation function over its worldsheet positions. Furthermore, $dV_{\textrm{CKG}}$ is the volume of the conformal Killing-group, which we parametrise as
\begin{align}
	dV_{\textrm{CKG}}=\frac{dz_1\,dy\,d\bar{y}}{(z_1-y)(z_1-\bar{y})(y-\bar{y})}\,.\label{VolumeCKG}
\end{align}
Since the vertex operator for $F^{\bar{S}'}$ consists of two parts corresponding to the NS and the R contribution, we can split the correlator (\ref{UnmixedDiag}), which appears as the integrand in (\ref{Corr11}), accordingly
\begin{align}
 D_{Y^{\dagger}a F^{\bar S'}}(z_1,z_2,y,\bar{y})= D_{Y^{\dagger}a F^{\bar S'}}^{\text{NS}}(z_1,z_2,y,\bar{y})+ D_{Y^{\dagger}a F^{\bar S'}}^{\text{R}}(z_1,z_2,y,\bar{y})\,.\label{Corr11NSR}
\end{align}
Starting with the NS contribution we find 
\begin{align}
D_{Y^{\dagger}a F^{\bar S'}}^{\text{NS}}=-\frac{g_6^2}{8\pi}\,Y_{\mu}^{\dag}a_{\nu}F_{\rho\sigma}^{\bar S'}\,&\left\{\left<\Psi(z_1)\bar\Psi(y)\right>\left<\psi^{\mu}(z_1)\psi^{\nu}(z_2)\psi^{\sigma}(y)\psi^{\rho}(y)\right>\left<e^{-\varphi(z_2)}e^{-\varphi(\bar y)}\right>-\right.\nonumber\\
 &\left.-\left<\Psi(z_1)\bar\Psi(\bar y)\right>\left<\psi^{\mu}(z_1)\psi^{\nu}(z_2)\psi^{\rho}(\bar y)\psi^{\sigma}(\bar y)\right>\left<e^{-\varphi(z_2)}e^{-\varphi(y)}\right>\right\}\,.\nonumber
\end{align}
Using the relations given in Appendix \ref{NoteConv}, we find
\begin{align}
D_{Y^{\dagger}a F^{\bar S'}}^{\text{NS}}&=-\frac{g_6^2}{8\pi}\,Y_{\mu}^{\dag}a_{\nu}F_{\rho\sigma}^{\bar S'}\,\times\,4\delta^{\mu\rho}\delta^{\nu\sigma}(z_1-y)^{-1}(z_1-\bar y)^{-1}(z_2-y)^{-1}(z_2-\bar y)^{-1}\,.\label{Corr11g}
\end{align}
The correlator $D_{Y^{\dagger}a F^{\bar S'}}^{\text{R}}$, which contains the R-part of the vertex (\ref{UbarVec}), can be computed in the same fashion and yields precisely the same answer. Furthermore, conformal symmetry still allows us to fix the position of one of the boundary insertions as well as the position of the bulk vertex in the correlator (\ref{Corr11NSR}), which we choose $z_1=\infty$ and $y=i$. Inserting (\ref{Corr11NSR}) and (\ref{Corr11g}) into (\ref{Corr11}) and integrating  $z_2$ over the real line, we obtain
\begin{align}
 \left<\left<V_{Y^{\dag}}\,V_a\,V_{F^{\bar S'}}\right>\right>_{\textrm{Disc}}=-4i\,\textrm{Tr}(Y_{\mu}^{\dag}a_{\nu}F_{\bar S'}^{\mu\nu})\,,\label{UnmixedTerm}
\end{align}
where we have used $2/(2\pi\alpha' g_6)^2=8\pi^2/ g_{YM}^{2}$ and restored the dimensionality of the fields in terms of $2\pi\alpha'$.\footnote{In \cite{Morales:1996bp,Antoniadis:2010iq} a different string background was studied, which involved the self-dual vector partner of the dilaton~$S$. However, repeating the computation of the above disc diagrams using this vector instead, leads to a vanishing result. This indicates that the background \cite{Morales:1996bp,Antoniadis:2010iq} does not give rise to the deformed ADHM action on a general $\Omega$-background.}

Applying the same techniques, we show that the diagram with boundary fermionic insertions \eqref{UnmixedFerm} vanishes identically:
\begin{align}
	\left<\left<V_{M}\,V_{M}\,V_{F^{\bar S'}}\right>\right>=\frac{8\pi^2}{g_{\textrm{YM}}^2}\int\frac{dz_1\, dz_2\, dy \,d\bar y}{dV_{\textrm{CKG}}}D_{MM F^{\bar S'}}(z_1,z_2,y,\bar{y})=0 \,.
\end{align}


\subsubsection*{$D_5-D_9$ disc diagrams}

Let us also discuss the contributions of mixed diagrams corresponding to the correlator \eqref{MixedDiag}. In this case, the disc amplitude can be computed in the same way as in the D5-D5 sector:
\begin{align}\label{MixedTerm}
 \left<\left<V_{\bar X^{\dag}}\,V_{\omega}\,V_{F^{\bar S'}}\right>\right>=\frac{8\pi^2}{g_{\textrm{YM}}^2}\int\frac{dz_1\, dz_2\, dy \,d\bar y}{dV_{\textrm{CKG}}}D_{\bar X^{\dagger}\omega F^{\bar S'}}(z_1,z_2,y,\bar{y}) \,.
 \end{align}
Going through similar steps as before, we find
 \begin{align}
 \left<\left<V_{\bar X^{\dag}}\,V_{\omega}\,V_{F^{\bar S'}}\right>\right>= \frac{i}{2}\,\textrm{Tr}(X^{\dag}_{\dot\alpha}\,\omega_{\dot\beta}\,(\bar\sigma^{\mu\nu})^{\dot\alpha\dot\beta}\,F^{\bar S'}_{\mu\nu})\,.\label{Answer55}
\end{align}
These diagrams are sufficient to compute the tree-level string effective action involving D5-instantons. In Section \ref{Sec:ADHM}, we consider the field theory limit and compare the result to the ADHM action on a general $\Omega$-background.
\subsubsection*{Picture Dependence of Unphysical Diagrams}\label{Sect:PictureDependence}
It is important to note that correlation functions involving auxiliary fields depend on the picture one uses for the vertex operators. This is to be expected since auxiliary fields are not BRST-closed and, hence, their use in string amplitudes can lead to ambiguous results. However, as proven in Appendix \ref{App:PictureIndependence}, the specific prescription for the pictures used in the previous section leads to the correct physical result. 

\begin{figure}[h!t]
\begin{center}
\parbox{8cm}{\epsfig{file=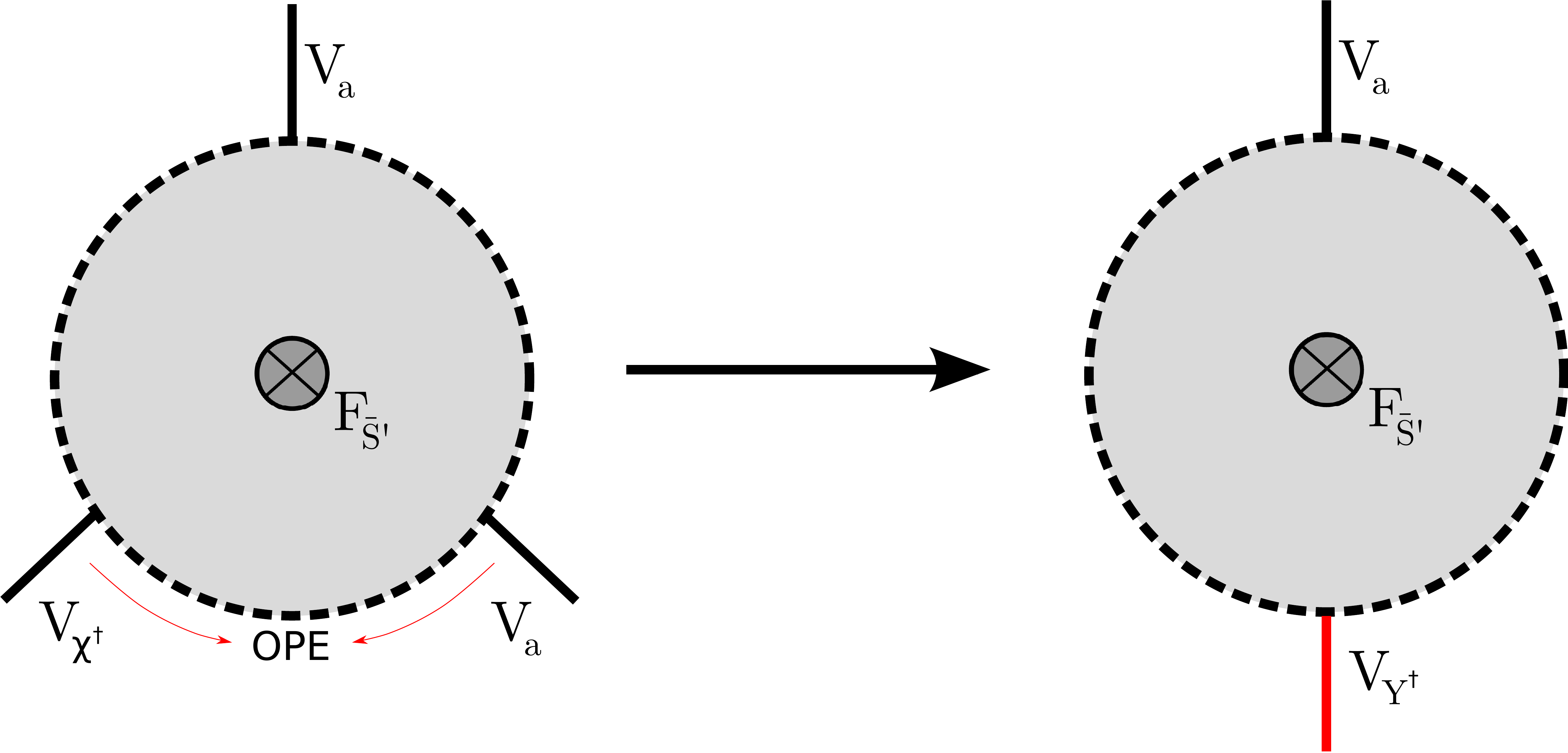,width=7.5cm}}
\end{center}
\hspace{1cm}\parbox{0.89\textwidth}{\caption{\small{\it Diagrammatic representation of the factorisation limit relating the correlator of physical operators with the one using an auxiliary field.}}\label{fig-factorisation}} 
\end{figure}

In particular, we repeat the calculation of  the relevant disc diagrams, which now involve four-point functions of physical  vertex operators only, and we recover the same result in an unambiguous fashion, independently of the choice of ghost picture, in accordance with the general arguments of \cite{Friedan:1985ey}. We then show that, only for a special choice of picture, the decoupling of the longitudinal mode of $F^{\bar S'}$ is manifest and the four-point correlator factorizes in such a way that allows one to effectively reinterpret the OPE of two physical vertices in terms of the auxiliary field used in the previous section (see~\cite{Atick:1987qy} for an earlier discussion of similar issues). This is pictured in Fig. \ref{fig-factorisation}. For other choices of pictures, it turns out that the decoupling of the longitudinal mode is no longer manifest as a factorization in any single channel and the correct result is only obtained by combining various different contributions with different factorization 
structures. This prevents us from reinterpreting the amplitude in terms of a simple three-point function involving an effective (auxiliary) vertex. Put differently, only with the picture prescription used in the previous section is the auxiliary field formalism applicable.

Therefore, working with auxiliary fields provides a simple framework that is justified \emph{only} through a calculation involving BRST-closed operators, as presented in Appendix \ref{App:PictureIndependence}.


\subsection{ADHM Action and Nekrasov Partition Function}\label{Sec:ADHM}

In this section we compare our results with the deformed ADHM action appearing in \cite{Nekrasov:2002qd}. The latter describes instantons in a gauge theory with $SU(N)$ gauge group on a general $\Omega$-background.  Our first step is to relate the states arising from the superstring construction described above with the physical gauge theory fields describing, in particular, the position and the size of the instantons. Our notation follows \cite{Atiyah:1978ri} (see \emph{e.g.} \cite{Dorey:2002ik} for a review). At the core of the (undeformed) ADHM construction lies a specific ansatz for the gauge connection. Requiring this ansatz to be a solution of the Yang-Mills equations of motion gives rise to a number of constraints that can be encoded in an action principle. To obtain a deformation of this ADHM action, we finally implement the $\Omega$ background as a particular space-time $U(1)^2$ rotation, with parameters $\epsilon_{\pm}$.

To be more specific, consider the following ansatz for the $SU(N)$ gauge connection 
\begin{align}\label{GaugeConnection}
&(A_{\mu})_{uv}=\bar U_{u}^a\partial_{\mu}U_{av}\,,&&\text{with} &&\bar{U}^a_uU_{av}=\delta_{uv}\,.
\end{align}
Here we have introduced the ADHM index $a=1,\ldots,2k+N$ and $u,v=1\ldots,N$, with $k$ being the instanton number\footnote{For notational simplicity, we mostly suppress the indices $u,v$ in what follows.} and $\bar{U}$ is the hermitian conjugate of $U$.
The ansatz \eqref{GaugeConnection} is a solution of the Yang-Mills equations if the corresponding field-strength is self-dual. To find a solution for the matrix $U$ which has this property, we first notice that the operator ${P_{a}}^b\equiv U_a\,\bar U^b$ is a projector preserving $U$, \emph{i.e.} $P^2=P$ and $PU=U$. The field strength of $A_{\mu}$ can be written in terms of $P$ as
\begin{equation}\label{FieldStrength}
 F_{\mu\nu}=\partial_{[\mu}\bar U^a({\delta_{a}}^b-{P_a}^b)\partial_{\nu]}U_b\,.
\end{equation}
One is thus led to write the ansatz $1\!\!1-P=\Delta\,f\,\bar\Delta$, where the $[N+2k]\times[2k]$ matrix $\Delta$ is called the  \emph{ADHM matrix}, while $f$ is an arbitrary Hermitian matrix. If we assume that $\Delta$ is linear in the space-time coordinates $x_{\mu}$
\begin{equation}
 \Delta_{ai\dot\alpha}=\mathfrak{a}_{ai\dot\alpha}+\mathfrak{b}^{\alpha}_{ai}\,x_{\alpha\dot\alpha}\,,
\end{equation}
and that the matrix $f$ can be `diagonalised' as
\begin{equation}
 {{f_{ij}}^{\dot\alpha}}_{\dot\beta}=f_{ij}{\delta^{\dot\alpha}}_{\dot\beta}\,,
\end{equation}
then the field strength \eqref{FieldStrength} is self-dual, being proportional to the matrix $\sigma_{\mu\nu}$, provided that the constraint
\begin{equation}\label{Constraint}
	f^{-1} 1\!\!1 =\bar\Delta\Delta \,,
\end{equation}
is satisfied. Here, the space-time coordinates are expanded in a spinor basis, \emph{i.e.} $x_{\alpha\dot\alpha}=x_{\mu}(\sigma^{\mu})_{\alpha\dot\alpha}$ and we have introduced the  instanton index $i=1,\ldots,k$. 
In this way, the construction of the ADHM solution is reduced to finding a consistent set of matrices $\mathfrak{a}$ and $\mathfrak{b}$. The symmetries of the problem allow us to write \cite{Dorey:2002ik}
\begin{align}
&\mathfrak{a}=\left(\begin{array}{c}
w \\
a' \end{array} \right)\,,&&\mathfrak{b}=\left(\begin{array}{c}
0_{[N]\times[2k]} \\
1\!\!1_{[2k]\times[2k]} \end{array} \right) \,,\label{ADHMansatz}
\end{align}
where $w$ and $a'$ are $[N]\times[2k]$ and $[2k]\times[2k]$ matrices respectively. In order to make contact with the degrees of freedom appearing in the string theory setup in Section~\ref{Sect:InstantonsSetup}, we identify
\begin{align}
&w=({\omega}_{\dot{\alpha}})_{ui}\,,&&a'=(a_{\alpha\dot{\alpha}})_{ij}\,.
\end{align}
The ADHM instanton action can finally be expressed as \cite{Nekrasov:2002qd}
\begin{align}\label{DeformedADHM}
 S_{\textrm{ADHM}}=-\textrm{Tr}&\left\{[\chi^{\dag},a_{\alpha\dot\beta}]\left([\chi,a^{\dot\beta\alpha}]+\epsilon_{-}(a\tau_3)^{\dot\beta\alpha}\right)-\chi^{\dag}\,\bar\omega_{\dot\alpha}\left(\omega^{\dot\alpha}\chi-\tilde a\,\omega^{\dot\alpha}\right)-\left(\chi\bar\omega_{\dot\alpha}-\bar\omega_{\dot\alpha}\,\tilde a\right)\omega^{\dot\alpha}\,\chi^{\dag}\right.\nonumber\\
   &\left.+\epsilon_+\left[\chi^{\dag},a_{\alpha\dot{\beta}}\right](\tau_3 a)^{\dot\beta\alpha}-\epsilon_+\,\bar{\omega}_{\dot{\alpha}}\,{(\tau_3)^{\dot{\alpha}}}_{\dot{\beta}}\,\chi^{\dag}\,\omega^{\dot{\beta}}\right\}\,.
\end{align}
where we only display the part relevant for our discussion. Here we have introduced the vev $\tilde a$ of the $\N=2$ vector multiplet that higgses the $SU(N)$ gauge group. The terms in the second line correspond to the $\epsilon_{+}$-dependent deformation of the ADHM action and which we want to compare to the effective couplings of $F^{\bar S'}$ to the ADHM moduli. To this end, we parametrize the vev of the $\bar S'$ field strength using the self-dual 't Hooft symbols (see equ. (\ref{tHooft})):
\begin{equation}
 F_{\mu\nu}^{\bar S'}\equiv\bar\eta_{\mu\nu}^{c}\,F_c^{\bar S'}\equiv\bar\eta_{\mu\nu}^{c}\,\delta_{3c}\,\frac{\epsilon_+}{2}\,.
\end{equation}
After integrating out the auxiliary fields, the contribution of the diagrams \eqref{UnmixedTerm} becomes  (see also \eqref{UnmixedPhys})
\begin{align}
 \left<\left<V_{Y^{\dag}}\,V_a\,V_{F^{\bar S'}}\right>\right>= -4i \,\textrm{Tr}\Bigr\{\left[\chi^{\dag},a_\mu\right]a_{\nu}\,F^{\mu\nu}_{\bar S'}\Bigr\}&=-\epsilon_+ \,\textrm{Tr}\Bigr\{\left[\chi^{\dag},a_{\alpha\dot\beta}\right]a^{\dot\gamma\alpha}\,{(\tau^{3})^{\dot\beta}}_{\dot\gamma}\Bigr\}\,.
\end{align}
Similarly, after integrating out the auxiliary fields, the contribution \eqref{MixedTerm} of the mixed diagrams can be recast as 
\begin{align}
 \left<\left<V_{\bar X^{\dag}}\,V_{\omega}\,V_{F^{\bar S'}}\right>\right>=\frac{i}{2}\,\textrm{Tr}\Bigr\{\bar\omega_{\dot\alpha}\,\chi^{\dag}\,\omega_{\dot\beta}(\bar\sigma^{\mu\nu})^{\dot\alpha\dot\beta}F_{\mu\nu}^{\bar S'}\Bigr\}=\epsilon_+\,\textrm{Tr}\Bigr\{\bar\omega_{\dot\alpha}\,\chi^{\dag}\,{(\tau_{3})^{\dot\alpha}}_{\dot\beta}\,\omega^{\dot\beta}\Bigr\}\,.
\end{align}
Therefore, the D5-instanton world-volume theory in our background gives a six-dimensional field theory on $K3\times T^2$ that contains the deformed ADHM couplings. In the limit where the world-volume of the D5-instantons becomes 
small (\emph{i.e.} of order $\alpha'$ or smaller), we can reduce the six-dimensional field theory to a zero-dimensional one which exactly reproduces the deformed ADHM action for the four-dimensional gauge theory.

The ADHM action is a key ingredient to compute the full Nekrasov partition function $Z^{\text{Nek}}$ of the supersymmetric gauge theory. It can be factorized in the following form:
\begin{align}
Z^{\text{Nek}}(\epsilon_+,\epsilon_-)=Z^{\text{Nek}}_{\text{pert}}(\epsilon_+,\epsilon_-)\, Z^{\text{Nek}}_{\text{n.p.}}(\epsilon_+,\epsilon_-)\,.\label{MarEq}
\end{align}
While the perturbative piece $Z^{\text{Nek}}_{\text{pert}}$ does not receive contributions beyond the one-loop order, the non-perturbative part $Z^{\text{Nek}}_{\text{n.p.}}$ is defined as a path integral over the instanton moduli space, with the integral measure given by the deformed ADHM action $S_{\text{ADHM}}$~\cite{Nekrasov:2002qd,Nekrasov:2003rj}.
\section{Interpretation and the Refined Topological String}\label{Sec:RefTop}

In this section we interpret our results in terms of recent developments in topological string theory. Inspired by the work of Nekrasov~\cite{Nekrasov:2002qd,Nekrasov:2003rj}, it has been realised that a one-parameter extension exists, known as the \emph{refined topological string}, such that its partition function at genus $g$, in the field theory limit $\mathcal{F}_{g,n}^{\textrm{f.t.}}$, reproduces the partition function of a gauge theory in the $\Omega$-background \cite{Nekrasov:2002qd,Nekrasov:2003rj,Losev:2003py,Iqbal:2003ix,Iqbal:2003zz}
\begin{equation}
	\log Z^{\textrm{Nek}}(\epsilon_{+},\epsilon_{-}) = \frac{1}{\epsilon_{-}^2-\epsilon_{+}^2} \sum_{g,n} \epsilon_{-}^{2g-2}\epsilon_{+}^{2n}\,\mathcal{F}_{g,n}^{\textrm{f.t.}}\,.
\end{equation}
 Several proposals for a description of the refined topological string have been presented in the literature \cite{Hollowood:2003cv,Dijkgraaf:2006um,Iqbal:2007ii,Lockhart:2012vp}. However, a convincing worldsheet description has turned out to be rather challenging \cite{Antoniadis:2010iq,Nakayama:2011be}.

Recently, we have put forward a promising proposal in \cite{Antoniadis:2013bja}, in terms of a particular class of higher derivative terms in the string effective action.  At the component level, it involves terms of the form
\begin{align}
\mathcal{I}_{g,n} = \int d^4x\,&\mathcal{F}_{g,n}\, R_{(-)\, \mu\nu\rho\tau} 
R_{(-)}^{\mu\nu\rho\tau}\, \left[F^G_{(-)\,\lambda\sigma} F^{G\,\lambda\sigma}_{(-)}\right]^{g-1}\,\left[F_{(+)\,\rho\sigma}F^{\rho\sigma}_{(+)}\right]^n
\label{EffectiveCoupling}\,,
\end{align}
for $g\geq1$ and $n\geq 0$. Here, $R_{(-)}$ denotes the anti-self-dual Riemann tensor, $F^G_{(-)}$ the anti-self-dual field strength tensor of the graviphoton and $F_{(+)}$ the self-dual field strength tensor of an additional vector multiplet gauge field. In Heterotic $\mathcal{N}=2$ compactifications on $K3\times T^2$, the latter is identified with the super-partner of the K\"ahler modulus of $T^2$, while in the dual Type I setting, it is mapped to the vector partner of the $\bar{S}'$ modulus. 

For $n=0$, the $\mathcal{I}_{g,n}$ in (\ref{EffectiveCoupling}) are BPS-saturated and were first discussed in \cite{Antoniadis:1993ze}. The $F_g=\mathcal{F}_{g,n=0}$ are exact at the $g$-loop level in Type II string theory compactified on an elliptically fibered Calabi-Yau threefold and compute the corresponding genus $g$ partition function of topological string theory \cite{Antoniadis:1993ze}. In the dual heterotic theory,  $F_g$ starts receiving contributions at the one-loop level~\cite{Antoniadis:1995zn} and, in the point particle limit, is related to the perturbative part of Nekrasov's partition function for a gauge theory on an $\Omega$-background with only one non-trivial deformation parameter. The latter is then identified with the topological string coupling. 

For $n>0$ the leading contribution to $\mathcal{F}_{g,n}$ is still given by a one-loop amplitude in the heterotic theory and was computed to all orders in $\alpha'$ in \cite{Antoniadis:2013bja}. The coupling functions $\mathcal{F}_{g,n}$ in (\ref{EffectiveCoupling}) can be compactly expressed in the form of a generating functional
\begin{align}\label{FullAmplitudeHet}
 \mathcal{F}\textrm(\epsilon_{-},\epsilon_{+})&=\sum_{g,n\geq0}\epsilon_-^{2g}\epsilon_+^{2n}\,\mathcal{F}_{g,n}\,.
\end{align}
In the point particle limit, the one-loop contribution to $\mathcal{F}(\epsilon_-,\epsilon_+)$ captures the perturbative part of the Nekrasov partition function (\ref{MarEq}) on a generic $\Omega$-background, whose deformation parameters are identified with the expansion parameters $\epsilon_\pm$ in (\ref{FullAmplitudeHet}). Thus, the $\mathcal{F}_{g,n}$ are  one-parameter extensions of the topological amplitudes $F_g$ which are (perturbatively) compatible with a refinement in the gauge theory limit.

The instanton computations performed in the previous sections are indeed evidence for this proposal. As we showed, also the non-perturbative contributions to the $\mathcal{F}_{g,n}$ in the point particle limit are compatible with the structure expected from gauge theory and capture precisely the non-perturbative part of the full Nekrasov partition function $Z^{\textrm{Nek}}_{\textrm{n.p.}}(\epsilon_{+},\epsilon_{-})$. This suggests that the couplings~(\ref{EffectiveCoupling}) indeed provide a string theoretic realisation of the full $\Omega$-background in gauge theory. This is precisely what one would expect from a worldsheet realisation of the refined topological string.


\section{Conclusions}\label{Sec:Conclusions}

In this paper we computed gauge theory instanton corrections to the class of higher derivative couplings, introduced in \cite{Antoniadis:2013bja}. These were obtained from tree level amplitudes in Type I string theory compactified on $K3\times T^2$, where we allowed for bulk insertions of anti-self-dual graviphotons as well as self-dual field strength tensors $F_{\bar{S}'}$ of the vector partner of $\bar{S}'$. In the field theory limit, these amplitudes reproduce correctly the fully $\Omega$-deformed ADHM action, \emph{i.e.} with both deformation parameters $\epsilon_{+},\epsilon_{-}\neq 0$. Using localisation, the resulting path interal over the instanton moduli can be computed \cite{Nekrasov:2002qd,Nekrasov:2003rj,Losev:2003py} yielding the non-perturbative gauge theory partition function.

Our work generalises the results of \cite{Billo:2006jm}, where the deformed ADHM action in the limit $\epsilon_+=0$ was reproduced from a pure graviphoton background, and extends the calculations of \cite{Ito:2010vx,Ito:2011cr} for $\epsilon_+\neq0$ by including NS-NS field strengths. Hence, building on our earlier results \cite{Antoniadis:2013bja}, we proved that the specific class of higher derivative string couplings (\ref{EffectiveCoupling}) correctly reproduces, in the point particle limit, the partition function of pure $\mathcal{N}=2$ gauge theory, \emph{both} perturbatively and non-perturbatively.

We would like to stress that $F_{\bar{S}'}$ is crucial for obtaining the correct $\Omega$-deformation of the ADHM action. For instance, we have checked that a background including self-dual field strength tensors of the vector partner of the heterotic dilaton~\cite{Antoniadis:2010iq}, does not lead to the full deformation of the ADHM action. It remains as an interesting question what such deformations mean from a gauge theory perspective.

As a further consequence, our findings also provide strong support to our proposal that the couplings considered in \cite{Antoniadis:2013bja} can be interpreted as a worldsheet description of the refined topological string. Indeed, these couplings generalise a class of BPS-saturated couplings $F_g$ discussed in \cite{Antoniadis:1993ze,Antoniadis:1995zn,Antoniadis:1996qg}, which capture the genus $g$ partition function of the `unrefined' topological string. The additional coupling to $F_{\bar{S}'}$ yields a one-parameter extension which  corresponds to a refinement in the field theory limit, as was discussed perturbatively in \cite{Antoniadis:2013bja}. Our current work indicates that this also carries over in the presence of instanton effects. Thus, these couplings satisfy a number of properties one would expect from a worldsheet realisation of the refined topological string. It would be very interesting to further study their properties.

Finally, it would be interesting to study the connection between our results and recent proposals for a realisation of the $\Omega$-background within string theory using the flux-trap background \cite{Hellerman:2011mv}.

\section*{Acknowledgements}
We thank C. Angelantonj, R. Blumenhagen, A. Klemm, N. Lambert, M. Mari\~no, N. Mekareeya and J.F. Morales for useful discussions. We would also like to thank M.~Bill\'o, M.~Frau, F.~Fucito and A.~Lerda  for pointing out their work to us and also J. Gomis for bringing \cite{Ito:2010vx,Ito:2011cr} to our attention. I.F. would like to thank the CERN Theory Division and A.Z.A. would like to thank the ICTP Trieste and Max-Planck-Institut f\"ur Physik in Munich for their warm hospitality during several stages of this work. This work was supported in part by the European Commission under the ERC Advanced Grant 226371 and the contract PITN-GA-2009- 237920.


\appendix

\section{Notations and useful results}\label{NoteConv}
In this appendix we summarise our conventions and notations.
\subsection{Indices}
Let us begin by discussing our conventions for various index structures. Raising and lowering of $SO(4)_{ST}$ sponsorial indices is achieved with the help of the epsilon-tensors $\epsilon^{12}=\epsilon_{12} = +1$, and $\epsilon^{\dot{1}\dot{2}}=\epsilon_{\dot{1}\dot{2}} = -1$, i.e.
\begin{align}
&\psi^\alpha = +\epsilon^{\alpha\beta}\psi_\beta \,,&&\psi_\alpha=-\epsilon_{\alpha\beta}\psi^\beta\,,&&\psi^{\dot\alpha} = -\epsilon^{\dot\alpha\dot\beta}\psi_{\dot\beta} \,,&&\psi_{\dot\alpha}=+\epsilon_{\dot\alpha\dot\beta}\psi^{\dot\beta}\,.
\end{align}
Furthermore we introduce the $\sigma$-matrices $(\sigma^\mu)_{\alpha\dot\alpha}$ and $(\bar\sigma^\mu)^{\dot\alpha\alpha}$ of $SO(4)_{ST}$
\begin{align}
	(\sigma^\mu)_{\alpha\dot\alpha} = (1\!\!1,-i\boldsymbol{\sigma})\,, && (\bar\sigma^\mu)^{\dot\alpha\alpha} = (1\!\!1,+i\boldsymbol{\sigma})\,,
\end{align}
which are related to one-another by raising and lowering of the spinor indices
\begin{align}
	(\sigma^\mu)^{\beta\dot\beta}\equiv\epsilon^{\beta\alpha}(\sigma^\mu)_{\alpha\dot\alpha}\epsilon^{\dot\alpha\dot\beta} = (\bar\sigma^\mu)^{\dot\beta\beta}\,, &&
	(\bar\sigma^\mu)_{\dot\beta\beta}\equiv \epsilon_{\dot\beta\dot\alpha}(\bar\sigma^\mu)^{\dot\alpha\alpha}\epsilon_{\alpha\beta}= (\sigma^\mu)_{\beta\dot\beta}\,.
\end{align}
Furthermore, we introduce the Lorentz generators $\sigma^{\mu\nu}, \bar\sigma^{\mu\nu}$ of $SO(4)_{ST}$
\begin{align}
{(\sigma_{\mu\nu})_\alpha}^\beta \equiv \frac{1}{2}{\bigr( \sigma_\mu\bar\sigma_\nu-\sigma_\nu\bar\sigma_\mu\bigr)_{\alpha}}^{\beta}\,,&&{(\bar\sigma_{\mu\nu})^{\dot\alpha}}_{\dot\beta} \equiv \frac{1}{2}{\bigr( \bar\sigma_\mu\sigma_\nu-\bar\sigma_\nu\sigma_\mu\bigr)^{\dot\alpha}}_{\dot\beta}\,,
\end{align}
which are symmetric in the spinorial indices $(\sigma_{\mu\nu})_{\alpha\beta}=+(\sigma_{\mu\nu})_{\beta\alpha}$ and $(\bar\sigma_{\mu\nu})_{\dot\alpha\dot\beta}=+(\sigma_{\mu\nu})_{\dot\beta\dot\alpha}$. Furthermore, these generators are (anti-)self-dual in the sense
\begin{align}
(\sigma^{\mu\nu})_{\alpha\beta} = +\frac{1}{2}\epsilon^{\mu\nu\rho\sigma} (\sigma_{\rho\sigma})_{\alpha\beta}\,, &&(\bar\sigma^{\mu\nu})_{\dot\alpha\dot\beta} = -\frac{1}{2}\epsilon^{\mu\nu\rho\sigma} (\bar\sigma_{\rho\sigma})_{\dot\alpha\dot\beta}\,.
\end{align}
Therefore, we can use them to define \emph{(anti-)self-dual} tensors respectively. In particular we write for the self-dual $F_{\mu\nu}^{(+)}$ and anti-self-dual $F_{\mu\nu}^{(-)}$ part of the field strength tensor of a given gauge field
\begin{align}
F^{(+)}_{\dot\alpha\dot\beta}\equiv (\bar\sigma^{\mu\nu})_{\dot\alpha\dot\beta}F^{(+)}_{\mu\nu}\,,&&F^{(-)}_{\alpha\beta}\equiv (\sigma^{\mu\nu})_{\alpha\beta} F^{(-)}_{\mu\nu}\,.\label{SelfDualDef}
\end{align}
with $F^{(\pm)}_{\mu\nu} = \mp \frac{1}{2} \epsilon_{\mu\nu\rho\sigma} (F^{(\pm)})^{\rho\sigma}$. Also, note the following identities:
\begin{align}
	&(\sigma^{\mu\nu})_{\alpha\beta}(\sigma^{\rho\sigma})^{\alpha\beta} = 2(\delta^{\mu\rho}\delta^{\nu\sigma}-\delta^{\mu\sigma}\delta^{\nu\rho}+\epsilon^{\mu\nu\rho\sigma})\,,\\
	&(\bar\sigma^{\mu\nu})_{\dot\alpha\dot\beta}(\bar\sigma^{\rho\sigma})^{\dot\alpha\dot\beta} = 2(\delta^{\mu\rho}\delta^{\nu\sigma}-\delta^{\mu\sigma}\delta^{\nu\rho}-\epsilon^{\mu\nu\rho\sigma})\,.
\end{align}
Using the above relations, one may invert (\ref{SelfDualDef}) to obtain
\begin{align}
	& F_{\mu\nu}^{(+)} = \frac{1}{8}(\bar\sigma_{\mu\nu})^{\dot\alpha\dot\beta}F^{(+)}_{\dot\alpha\dot\beta}\,,&&
	& F_{\mu\nu}^{(-)} = \frac{1}{8}(\sigma_{\mu\nu})^{\alpha\beta}F^{(-)}_{\alpha\beta}\,.
\end{align}
Finally, we define the (anti-)self-dual 't Hooft symbols by decomposing the sigma-matrices:
\begin{align}
 {(\bar\sigma_{\mu\nu})^{\dot\alpha}}_{\dot\beta}\equiv i\,\bar\eta_{\mu\nu}^{c}\,{(\tau_c)^{\dot\alpha}}_{\dot\beta}\,,&&{(\sigma_{\mu\nu})_{\alpha}}^{\beta}\equiv i\,\eta_{\mu\nu}^{c}\,{(\tau_c)_{\alpha}}^{\beta}\,.\label{tHooft}
\end{align}
The notation we use closely follows \cite{Billo:2006jm}. Self-dual spin-fields of $SO(4)_{ST}$ are denoted $S_{\dot\alpha}$ and the anti-self-dual ones $S_{\alpha}$. In this notation, the graviphoton field $G$ is anti-self-dual and the $\bar S'$ self-dual.

\subsection{Correlators}

We summarize here the main correlation functions needed for the calculation of the disc diagrams encountered in the text:
{\allowdisplaybreaks 
\begin{align}
\left<\Psi(z_1)\bar\Psi(z_2)\right> &= z_{12}^{-1}\,,\nonumber\\[6pt]
\left<\bar{\Delta}(z_1)\Delta(z_2)\right>&=-z_{12}^{-1/2}\,,\nonumber\\[6pt]
\left<e^{-\varphi(z_1)}e^{-\tfrac{1}{2}\varphi(z_2)}e^{-\tfrac{1}{2}\varphi(z_3)}\right>&=z_{12}^{-1/2}z_{13}^{-1/2}z_{23}^{-1/4}\,,\nonumber\\[6pt]
\left<\Psi(z_1)S^{\hat A}(z_2)S^{\hat B}(z_3)\right>&=-i\,\epsilon^{\hat A\hat B}\,z_{12}^{-1/2}z_{13}^{-1/2}z_{23}^{-1/4} \,,\nonumber\\[6pt]
\left<\psi^{\mu}(z_1)\psi^{\nu}(z_2)\psi^{\rho}(z_3)\psi^{\sigma}(z_4)\right>&=\delta^{\mu\nu}\delta^{\rho\sigma}\,z_{12}^{-1}z_{34}^{-1}-\delta^{\mu\rho}\delta^{\nu\sigma}\,z_{13}^{-1}z_{24}^{-1}+\delta^{\mu\sigma}\delta^{\nu\rho}\,z_{14}^{-1}z_{23}^{-1}\,,\nonumber\\[6pt]
\left<S^{\dot\alpha}(z_1)S^{\dot\beta}(z_2)S^{\dot\gamma}(z_3)S^{\dot\delta}(z_4)\right>&=\frac{\epsilon^{\dot\alpha\dot\beta}\epsilon^{\dot\gamma\dot\delta}\,z_{14}z_{23}-\epsilon^{\dot\alpha\dot\delta}\epsilon^{\dot\beta\dot\gamma}\,z_{12}z_{34}}{(z_{12}z_{13}z_{14}z_{23}z_{24}z_{34})^{1/2}}\,,\nonumber\\[6pt]
\left<\psi^{\mu}(z_1)\psi^{\nu}(z_2)S^{\dot\alpha}(z_3)S^{\dot\beta}(z_4)\right>&=-\frac{z_{13}z_{24}+z_{23}z_{14}}{2\,z_{12}(z_{13}z_{14}z_{23}z_{24}z_{34})^{1/2}}\delta^{\mu\nu}\epsilon^{\dot\alpha\dot\beta}-\frac{z_{34}^{1/2}}{2\,(z_{13}z_{14}z_{23}z_{24})^{1/2}}(\bar\sigma^{\mu\nu})^{\dot\alpha\dot\beta}\,.\label{Correlators}
\end{align}
}
Here we have introduced the shorthand notation $z_{ij}=z_i-z_j$.

\section{Picture Independence of Disc Amplitudes}\label{App:PictureIndependence}
In this appendix we repeat the disc computations performed in Section~\ref{Sect:DiskDiagrams} making  use of physical vertex operators only.

\subsection{Picture Dependence of Disc Diagrams}
The auxiliary fields introduced in (\ref{Auxiliary1})--(\ref{Auxiliary3}) are not physical and, therefore, they do not lie in the BRST cohomology of the theory. Inserting such fields as vertex operators into string scattering amplitudes typically leads to an unphysical dependence on the ghost-picture \cite{Friedan:1985ey}, thus rendering the result ambiguous. 
To illustrate this point, consider, for instance, the unmixed diagram \eqref{Corr11} and focus, for simplicity, on the NS-NS part of bulk vertex operator $V_{F^{\bar S'}}$ which we now insert in a different picture:
\begin{equation}
 V_{F^{\bar S'}}^{\textrm{NS}}~\propto~ P_\rho\epsilon_\sigma(e^{-\varphi(\bar y)}\psi^{\sigma}(y)\bar\Psi(y)\psi^{\rho}(\bar y)+e^{-\varphi(y)}\psi^{\rho}(y)\psi^{\sigma}(\bar y)\bar\Psi(\bar y))\,.
\end{equation}
The correlator can be calculated as before:
\begin{align}
 D_{Y^{\dagger}\,a\, F^{\bar S'}}^{\textrm{NS}}(z_1,z_2,y,\bar{y})~\propto~ \frac{Y^\dagger_\mu\,a_\nu\, P^\mu\,\epsilon^\nu}{(z_1-\bar z)^2|z_2-z|^2}-\frac{Y^\dagger_\mu\,a_\nu\, P^\nu\,\epsilon^\mu}{(z_2-\bar z)^2|z_1-z|^2}\,.\label{PicDepDiag}
\end{align}
Including the factor \eqref{VolumeCKG}, setting $z_1\rightarrow\infty$, $y\rightarrow i$ and integrating over $z_2$ then yields a different result than (\ref{Answer55}):
\begin{align}
 \left<\left<V_{Y^{\dag}}\,V_{a}\,V_{F^{\bar S'}}\right>\right>_{\textrm{NS}}~\propto~ Y^\dagger_\mu\,a_\nu\, P^\mu\,\epsilon^\nu\,.
\end{align}
Notice that only the first term in \eqref{PicDepDiag} gives a non-vanishing contribution. The result indicates that the longitudinal modes do not decouple and is due to the fact that the auxiliary field vertex operator is not annihilated by the string BRST operator.

\subsection{Physical Amplitudes}
In order to obtain an unambiguous answer, we calculate below the disc amplitudes without auxiliary fields and show how the results of Section~\ref{Sect:DiskDiagrams} can be recovered in an unambiguous fashion.

A crucial requirement for all our scattering amplitudes is that the external fields involved be physical, \emph{i.e.} that they be annihilated by the BRST operator
$Q=\oint J_{\text{BRST}}$. The BRST-current can be written in the form 
\begin{equation}
J_{\text{BRST}}= c(T_{\text{matter}}+T_{\text{superghost}})+c\partial c\,b+ \gamma\,T^F_{\text{matter}} -\gamma^2\,b \,,\label{BRSTcurrent}
\end{equation}
where the energy momentum tensors of the matter and superghost sectors, as well as their fermionic partners, take the form
\begin{alignat}{3}
&T_{\text{matter}} &=&\, -\frac{1}{2} \partial X^I \partial X_I - \frac{1}{2} \psi^I \partial\psi_I \,,\nonumber\\
&T_{\text{superghost}}&=&\,-\frac{1}{2} \gamma\,\partial \beta -\frac{3}{2} \partial \gamma\,\beta \,,\nonumber\\
&T^F_{\text{matter}} &=&\, i\psi^I \partial X_I\,.
\end{alignat}
Here we are using the following OPEs for the worldsheet and ghost fields:
\begin{align}
&\partial X^I(x) \partial X^J(y) = -\frac{\delta^{IJ}}{(x-y)^2}\,,&&\psi^I(x) \psi^J(y)=\frac{\delta^{IJ}}{x-y}\,,\nonumber\\
&\beta(x) \gamma(y) = -\frac{1}{x-y}\,,&&b(x) c(y) =\frac{1}{x-y}\,.\nonumber
\end{align}
As we show below, the physical amplitudes take the form of contact terms $p_i.p_j/p_i.p_j$ where $p_i$ are the momenta of the various vertex operator insertions. These contact terms give non-trivial results in the limit $p_i\to 0$. To be able to compute them in a well-defined manner, we keep $p_i$ generic in all intermediate steps, which also acts as a regularisation of the worldsheet integrals, and take the limit only at the end of the calculation. However, due to the nature of our vertex insertions, we cannot switch on momenta in an arbitrary fashion: since the four-dimensional space-time corresponds to directions with Dirichlet boundary conditions for the D5-instantons, none of the ADHM moduli can carry momenta along $X^\mu$. Similarly, the $\bar{S}'$-vector insertions cannot carry momenta along $T^2$ once we impose BRST invariance (\emph{i.e.} transversality and decoupling of longitudinal modes). As a way out, we take all vertices to carry momenta along the $K3$ directions and complexify them if 
necessary, to make all integrals well-defined. In fact, technically, we first replace $K3$ by $\mathbb{R}^4$ and compute an effective action term on the D-instanton world-volume $T^2 \times \mathbb{R}^4$. Since the 
relevant fields that appear in these couplings  survive the orbifold projection (or more generally on a smooth $K3$ manifold they give rise to zero-modes), the corresponding couplings exist also in the case where $\mathbb{R}^4$ is replaced by $K3$.

\begin{figure}[h!t]
\begin{center}
\parbox{4.2cm}{\epsfig{file=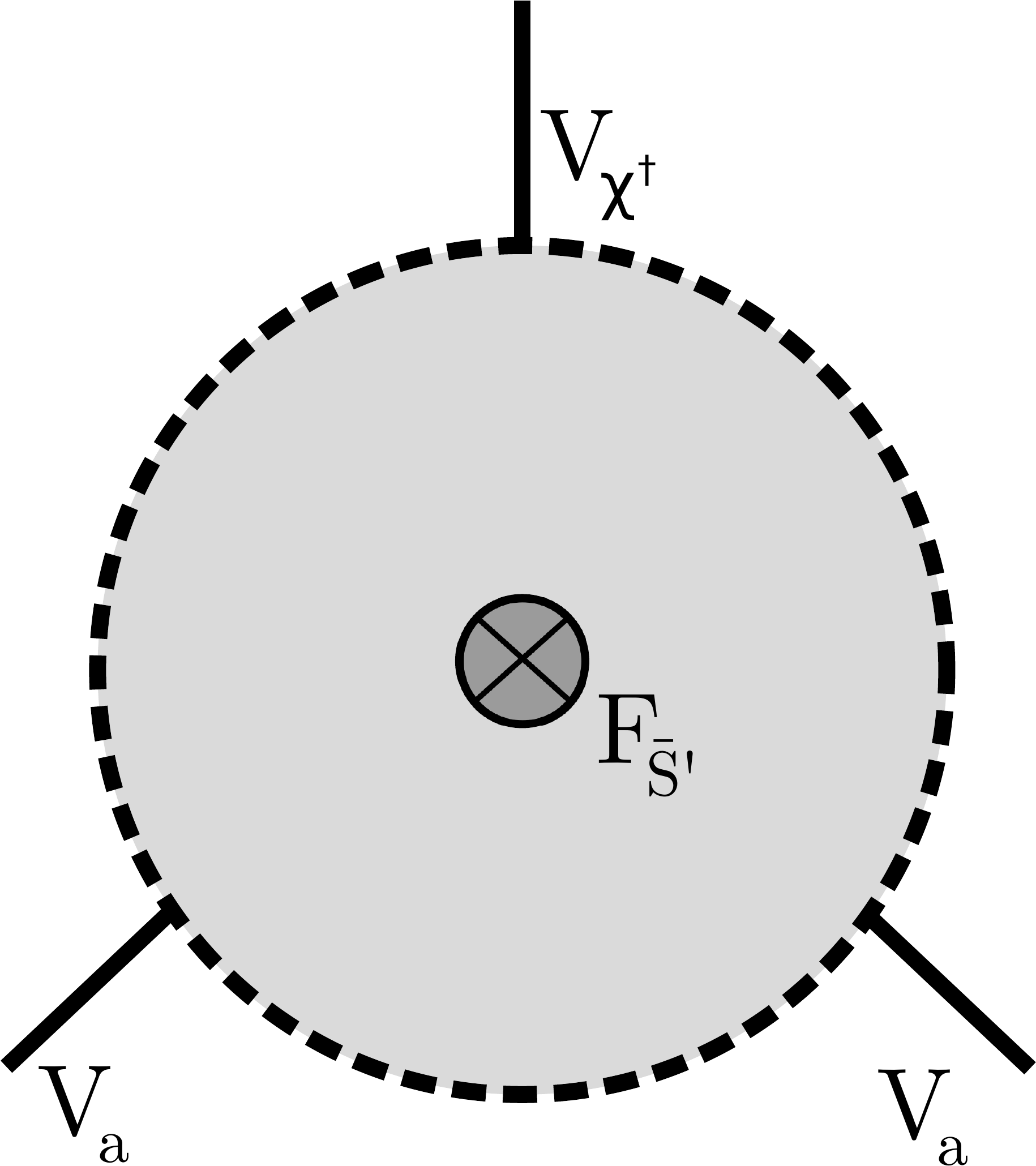,width=4.2cm}}
\end{center}
\hspace{1cm}\parbox{0.89\textwidth}{\caption{\small{\it Four-point disc diagram with bulk-insertion of the $\bar{S}'$ field strength tensor and  three boundary insertions stemming from the 5-5 sector of the string setup.}}\label{fig-fourpt}}
\end{figure}
We consider the physical four-point amplitude depicted in Fig. \ref{fig-fourpt}, which is of the form
\begin{equation}
\mathcal{D}_{aa\chi^{\dag} F^{\bar S'}}=\mathcal{D}_{aa\chi^{\dag} F^{\bar S'}}^{\textrm{NS}}+\mathcal{D}_{aa\chi^{\dag} F^{\bar S'}}^{\textrm{R}}\label{4ptFct}\,,
\end{equation}
with
\begin{align}
 \mathcal{D}_{aa\chi^{\dag} F^{\bar S'}}^{\textrm{NS}}&=\langle V^{(0)}_{a}(x_1) V_{a}^{(-1)}(x_2) V_{\chi^{\dag}}^{(0)}(x_3) V^{(-1,-1)}_{F^{\bar{S}'}}(z,\bar{z}) V_{\text{PCO}}(y) \rangle\label{4ptFctNS}\,,\\
 \mathcal{D}_{aa\chi^{\dag} F^{\bar S'}}^{\textrm{R}}&=\langle V^{(0)}_{a}(x_1) V_{a}^{(-1)}(x_2) V_{\chi^{\dag}}^{(0)}(x_3) V^{(-\frac{1}{2},-\frac{1}{2})}_{F^{\bar{S}'}}(z,\bar{z})\rangle\label{4ptFctR}\,.
\end{align}
Here  $V^{(-1,-1)}_{F^{\bar{S}'}}$ is the NS-NS part of the closed string $F^{\bar{S}'}$ field strength tensor introduced in (\ref{UbarVec}), $V^{(-\frac{1}{2},-\frac{1}{2})}_{F^{\bar{S}'}}$ is its R-R part and $V_{\text{PCO}}$ is the picture changing operator.\footnote{For the sake of clarity, we have explicitly denoted the ghost-picture of every vertex operator.} Upon bosonizing the $(\beta,\gamma)$ ghost system, introduced in (\ref{BRSTcurrent}), in terms of anti-commuting fields $(\eta, \xi)$ of dimension $(1,0)$, as well as a scalar $\phi$ with background charge $T_{\phi}=-1/2 ((\partial \phi)^2+ 2 \partial^2 \phi)$,
\begin{align}
&\beta= e^{-\varphi} \partial \xi\,,&&\gamma= e^{\varphi} \eta\,,
\end{align}
we can write $V_{\text{PCO}}= Q \xi$. In (\ref{4ptFct}) we have kept the PCO insertion at a fixed position $y$, even though the final result should not depend on $y$ \cite{Friedan:1985ey}. Setting $y$ to $z$, $\bar{z}$ or $x_2$, converts the ghost picture of the corresponding vertex operators to \footnote{In (\ref{4ptFct}) an insertion of $\xi$ at an arbitrary position is understood, in order to soak up the $\xi$ zero mode.}
\begin{align}
\mathcal{D}_{aa\chi^{\dag} F^{\bar S'}}^{\textrm{NS}}\big|_{y=z}&=\langle V^{(0)}_{a}(x_1) V_{a}^{(-1)}(x_2) V_{\chi^{\dag}}^{(0)}(x_3) V^{(0,-1)}_{F^{\bar{S}'}}(z,\bar{z}) \rangle\,,\label{4ptyz}\\
\mathcal{D}_{aa\chi^{\dag} F^{\bar S'}}^{\textrm{NS}}\big|_{y=\bar{z}}&=\langle V^{(0)}_{a}(x_1) V_{a}^{(-1)}(x_2) V_{\chi^{\dag}}^{(0)}(x_3) V^{(-1,0)}_{F^{\bar{S}'}}(z,\bar{z})\rangle\,,\label{4ptybz}\\
\mathcal{D}_{aa\chi^{\dag} F^{\bar S'}}^{\textrm{NS}}\big|_{y=x_2}&=\langle V^{(0)}_{a}(x_1) V_{a}^{(0)}(x_2) V_{\chi^{\dag}}^{(0)}(x_3) V^{(-1,-1)}_{F^{\bar{S}'}}(z,\bar{z}) \rangle\,,\label{4ptyx}
\end{align}
respectively.
Using the doubling trick, we can convert the disc into the full plane with a $\mb Z_2$-involution, and the four-point amplitude \eqref{4ptFct} becomes a five-point function with  vertices at $(x_1,x_2,x_3,z,\bar{z})$. Here we split
\begin{equation}
 V^{(-1,-1)}_{F^{\bar{S}'}}(z,\bar{z})=V_{F^{\bar{S}'}}(z)\,V_{F^{\bar{S}'}}(\bar{z})\,,
\end{equation}
where the left-right symmetrization is implicit. $SL(2,\mb R)$ invariance implies that we can fix three real positions, which is related to the existence of three $c$-ghost zero modes on the sphere. The latter are soaked up by attaching $c$ to three dimension one vertices such that the resulting operators are BRST closed.  The dimension of these vertices becomes zero and they remain unintegrated. Since the last two terms in (\ref{BRSTcurrent}) annihilate any operator in the $(-1)$-picture, any physical operator with dimension one and negative ghost picture becomes BRST invariant in this manner.\footnote{Indeed, the first two terms in (\ref{BRSTcurrent}) combined together annihilate $c V$ for any $V$ corresponding to a dimension one Virasoro primary operator, irrespective of the ghost picture of $V$.} However, when $c$ is attached to a vertex $V^{(0)}$ in the zero-picture, the last two terms in (\ref{BRSTcurrent}) do not annihilate $c V^{(0)}$. In this case, the correct  dimension zero BRST invariant 
combination is $c V^{(0)}+ \gamma V^{(-1)}$. Therefore, for simplicity, we choose the zero-picture vertices to be of dimension one (such that their positions are integrated),  and all the $(-1)$-picture vertices to be of dimension zero (such that their positions remain unintegrated).

Let us first consider the NS-NS contributions \eqref{4ptFctNS}, for which the vertex operators are
\begin{eqnarray}
V_{a}(x_1)&=& g_6\,a_{\mu}(\partial X^\mu-2 i  p_1\cdot\chi \,\psi^{\mu})e^{2 i p_1\cdot Y}(x_1)\,,\label{Va}\\
V_{a}(x_2)&=& g_6\,a_{\nu}\, c e^{-\varphi}\psi^{\nu}e^{2 i p_2\cdot Y}(x_2)\,,\\
V_{\chi^{\dag}}(x_3)&=&\frac{\chi^{\dag}}{\sqrt{2}}\, (\partial Z-2 i p_3\cdot\chi \,\Psi)e^{2 i p_3\cdot Y}(x_3)\label{Vchi}\,,\\
V_{F^{\bar{S}'}}(z)&=&  c e^{-\varphi}\bar{\Psi}e^{i(P_\mu X^\mu+ P\cdot Y)}(z)\,,\\
V_{F^{\bar{S}'}}(\bar{z})&=&-\frac{i\epsilon_{\lambda}}{8\pi\sqrt{2}}\,  c e^{-\varphi}\psi^{\lambda}e^{i(-P_\mu X^\mu+ P\cdot Y)}(\bar{z})\,,
\end{eqnarray}
with $F_{\mu\nu}^{\bar{S}'}\equiv \epsilon_{[\mu}P_{\nu]}$, and the only relevant terms in $V_{\text{PCO}}$ are (since the total background charge of the superghost is $-2$)
\begin{equation}
e^{\varphi}\, T^F_{\textrm{matter}}(y)= ie^{\varphi}(\psi^\mu \partial X_\mu+\Psi \partial \bar{Z}+\bar{\Psi} \partial{Z}+\chi^i \partial Y^i)(y)\,.
\end{equation}
Here, $Y^i\in \{X^6,X^7,X^8,X^9\}$ parametrize the internal $\mathbb{R}^4$ (which we eventually replace by $K3$). The momenta $p_i$ are along these directions, while the momentum of $V_{F^{\bar{S}'}}$ is written as $(P_\mu, P)$, where $P_\mu$ is the space-time part and $P$ is along the $Y^i$ directions. Note that after using the doubling trick, the Neumann directions $( Z, \bar{Z},Y^i)$ are mapped onto themselves, whereas the Dirichlet ones pick an additional minus sign $X^\mu \rightarrow -X^{\mu}$. This is consistent with the fact that the momenta along Neumann directions are conserved $\sum_i p_i + P =0$, which follows from integrating the zero modes of $y$. On the other hand, integrating over the zero modes of the Dirchlet directions $X^\mu$ does not give any conservation law for  $P_\mu$.

The three open string vertices contain Chan-Paton labels which need to be suitably ordered. For instance, if we are interested in computing the term $\textrm{Tr}(a_{\mu}\,a_{\nu}\,\chi^{\dag})$, the range of integration is the following:
\begin{equation}\label{IntRange}
\begin{cases}
 \textrm{for } x_1>x_2\,, & x_3\in ]x_2,x_1[\,,\\
 \textrm{for } x_2>x_1\,, & x_3\in]-\infty,x_1[\cup]x_2,\infty[\,.
\end{cases}
\end{equation}
For the other inequivalent ordering $\textrm{Tr}(a_{\mu}\,\chi^{\dag}\,a_{\nu})$, the range of the $x_3$-integration is opposite. It is easy to show that the sum of these two orderings vanishes so that the amplitude is of the form $\textrm{Tr}(a_{\mu}[\chi^{\dag}, a_{\nu}])$.

For definiteness, let us focus on the term $\textrm{Tr}(a_{\mu}\,a_{\nu}\,\chi^{\dag})$. The contraction of $\varphi$ and $c$ and the contraction of the exponentials in momenta yield
\begin{eqnarray}
A_0&=&\left\{-\frac{ig_6^2}{16\pi}\,\textrm{Tr}(a_{\mu}\,a_{\nu}\,\chi^{\dag})\epsilon_{\lambda}\right\}|y-z|^2(y-x_2)\nonumber\\
&\times&\prod_{1\leq i<j\leq3}(x_i-x_j)^{4 p_i\cdot p_j}\prod_{k=1}^{3}|x_k-z|^{4 p_k\cdot P}(z-\bar{z})^{- P_\mu P^\mu+ P_i P_i}\,.
\end{eqnarray}
This is a common factor that multiplies each of the remaining contractions. Now let us consider the contribution of $\partial{Z}(x_3)$ to the amplitude. This must contract with $\partial \bar{Z}(y)$ in $V_{\textrm{PCO}}$ and then $\Psi(y)$ contracts with $\bar\Psi(z)$. Then $\psi^{\lambda}(\bar{z})$ necessarily contracts with $\psi^\nu(x_2)$ and from $x_1$ only $\partial X^{\mu}(x_1)$ can contribute. The result is
\begin{equation}
A_1= \frac{i \delta^{\nu \lambda} P^\mu (z-\bar{z})}{(y-x_3)^2 (y-z)(x_2-\bar{z})|x_1-z|^2}\,.
\end{equation}
Next consider the contribution of the second term in \eqref{Vchi}. Here, there are two separate contributions.  If $p_3\cdot\chi(x_3)$ contracts with $p_1\cdot\chi(x_1)$, then $\psi^\mu(x_1), \psi^{\nu}(x_2), \psi^{\lambda}(\bar{z})$ and a space-time fermion $\psi^{\sigma}(y)$ from the picture changing operator must contract, leaving $\partial X^\sigma(y)$ which can only contract with the momentum parts of vertices at $z$ and $\bar{z}$ resulting in a term proprtional to $P_\sigma$. Notice that the term arising from the contraction of $\psi^\mu$ with $\psi^{\nu}$ is killed by the transversality condition (a necessary condition for the operator to be in the kernel of $Q$). The total result is
\begin{equation}
A_2= \frac{4 i p_1\cdot p_3\  (z-\bar{z})}{(x_3-z)(x_3-x_1)|y-z|^2}\left[\frac{\delta^{\nu \lambda} P^\mu}{(x_2-\bar{z})(y-x_1)}-\frac{\delta^{\mu \lambda} P^\nu}{(x_1-\bar{z})(y-x_2)}\right]\,.
\end{equation}
On the other hand, if $p_3\cdot\chi(x_3)$ contracts with $\chi(y)$ in $V_{\textrm{PCO}}$, then $\partial Y(y)$ contracts with momentum dependent parts of the vertices. Thus, $\psi^{\lambda}(\bar{z})$ must contract with $\psi^\nu(x_2)$ and only $\partial X^\mu$ at $x_1$ can contribute and one obtains
\begin{equation}
A_3=\frac{4 i\delta^{\nu \lambda} P^\mu (z-\bar{z})}{(x_3-z)(y-x_3)(x_2-\bar{z})|x_1-z|^2}\left[\frac{p_3\cdot p_1}{y-x_1}+\frac{p_3\cdot p_2}{y-x_2}+\frac{p_3\cdot P}{2(y-z)}+
\frac{p_3\cdot P}{2(y-\bar{z})}\right]\,.
\end{equation}
The total correlation function is thus 
\begin{equation}
 \mathcal{D}_{aa\chi F^{\bar S'}}^{\textrm{NS}}=A_0(A_1+A_2+A_3)\,,
\end{equation}
which must be integrated over $x_1$ and $x_3$. Note that all the terms in $A_1$, $A_2$ and $A_3$ come with one power of space time momentum $P^\mu$ which is exactly what is required to obtain a coupling to the field strength of the closed string gauge field. However, both $A_2$ and $A_3$ are quadratic in the momenta along the $Y^i$ directions and  they can only contribute to the amplitude in the zero-momentum limit if the integration over $x_1$ and $x_3$ gives a pole of the form $1/(p_a\cdotp p_b)$. Clearly, $A_0\cdotp A_3$ cannot provide such a pole (we are assuming a generic value of $y$ in the 
complex plane i.e. $\textrm{Im}(y)\neq 0$). On the other hand, the integral over $x_3$ for  $A_0\cdotp A_2$ gives a pole of the form $1/(p_1\cdotp p_3)$. Performing the  $x_3$ integral in both the regions \eqref{IntRange} yields precisely the same result, hence, the $x_1$ integral over the entire real line reads
\begin{equation}
A_0 A_2 = \frac{g_6^2}{8\pi}\,\textrm{Tr}\left[a_{\mu}\,a_{\nu}\,\chi^{\dag}\right]\int_{-\infty}^{\infty} dx_1 \frac{(z-\bar{z})}{|x_1-z|^2}\epsilon_{\lambda}\left[\frac{(x_1-\bar{z})(y-x_2)}{(x_2-\bar{z})(y-x_1)}P^{\mu}\delta^{\nu\lambda}-P^{\nu}\delta^{\mu\lambda}\right]\,,
\label{A0A2}
\end{equation}
where we have set all the momenta along the $Y^i$ directions to zero since there are no singularities in the remaining $x_1$ integral. Notice that $A_0\cdotp A_2$ alone does not lead to a gauge invariant answer. As for $A_0 A_1$ term the $x_3$ and $x_1$ integrals have no singularities and therefore momenta along $Y$ directions can be set to zero. The resulting $x_3$ integral for both regions \eqref{IntRange} gives precisely the same result:
\begin{equation}
A_0 A_1 = -\frac{g_6^2}{8\pi}\,\textrm{Tr}\left[a_{\mu}\,a_{\nu}\,\chi^{\dag}\right]\int_{-\infty}^{\infty} dx_1 \frac{(z-\bar{z})}{|x_1-z|^2}\epsilon_{\lambda}\frac{(y-\bar{z})(x_1-x_2)}{(x_2-\bar{z})(y-x_1)}P^{\mu}\delta^{\nu\lambda}\,.\label{A0A1}
\end{equation}
Adding the two terms \eqref{A0A2} and \eqref{A0A1}, we see that the result is gauge invariant. Performing the $x_1$ integration yields\footnote{Notice the additional factor of 2 due to the left-right symmetrization in the closed string vertex.}:
\begin{equation}\label{Phys4ptNS}
 \left<\left<V_{a}\,V_{a}\,V_{\chi}\,V_{F^{\bar S'}}^{\textrm{NS}}\right>\right>=-2i\,\textrm{Tr}\left[a_{\mu}\,a_{\nu}\,\chi^{\dag}\right]\epsilon_{\lambda}(P^{\mu}\delta^{\nu\lambda}-P^{\nu}\delta^{\mu\lambda})\,.
\end{equation}

Finally, let us consider the R-R contributions \eqref{4ptFctR}. The vertex operators are the same as above, except for the $\bar S'$-vector part which is given by
\begin{align}
V^{(-\frac{1}{2},-\frac{1}{2})}_{F^{\bar{S}'}}(z,\bar{z})=& P^\rho\,c\,e^{-\frac{\varphi}{2}}\,S_{\dot{\alpha}}\,S^{\hat A}\,e^{i(P\cdotp X+P\cdotp Y)}(z)\nonumber\\
&\times\epsilon_{\hat A\hat B}{(\bar{\sigma}^{\rho \lambda})^{\dot{\alpha}}}_{\dot{\beta}}\, 
c\,e^{-\frac{\varphi}{2}}\,S^{\dot{\beta}}\,S^{\hat B}\,e^{i(-P\cdotp X+P\cdotp Y)}(\bar{z})\,.
\end{align}
Since the total superghost charge of the vertices is $-2$, there is no need for a picture changing operator. The total charge in the torus plane implies that only $p_3\cdot\chi\,\Psi(x_3)$ in \eqref{Vchi} contributes so that only $p_1\cdot\chi\,\psi^{\mu}(x_1)$ in \eqref{Va} contributes. This term is proprtional to $p_1\cdotp p_3$. Once again the integral over $x_3$ gives a pole $1/(p_1\cdotp p_3)$ in the channel $x_3 \rightarrow x_1$. Performing the integrals over $x_1$ and $x_3$ as above leads to the same result:
\begin{equation}
 \left<\left<V_{a}\,V_{a}\,V_{\chi}\,V_{F^{\bar S'}}^{\textrm{NS}}\right>\right>=\left<\left<V_{a}\,V_{a}\,V_{\chi}\,V_{F^{\bar S'}}^{\textrm{R}}\right>\right>\,.
\end{equation}
In order to compare with the result obtained with the auxiliary vertices, we sum over the inequivalent orderings of the open vertex operators, which yields
\begin{equation}\label{UnmixedPhys}
 \left<\left<V_{a}\,V_{a}\,V_{\chi}\,V_{F^{\bar S'}}\right>\right>=-4i\,\textrm{Tr}\left[\chi^{\dag},a_{\mu}\right]a_{\nu}\,F_{\bar S'}^{\mu\nu}\,.
\end{equation}
Consequently, the use of auxiliary fields with the picture prescription of Section \ref{Sect:DiskDiagrams} leads to the correct physical results. 

\subsection{Channel Factorization and Auxiliary Fields}
\begin{figure}[h!t]
\begin{center}
\parbox{13cm}{\epsfig{file=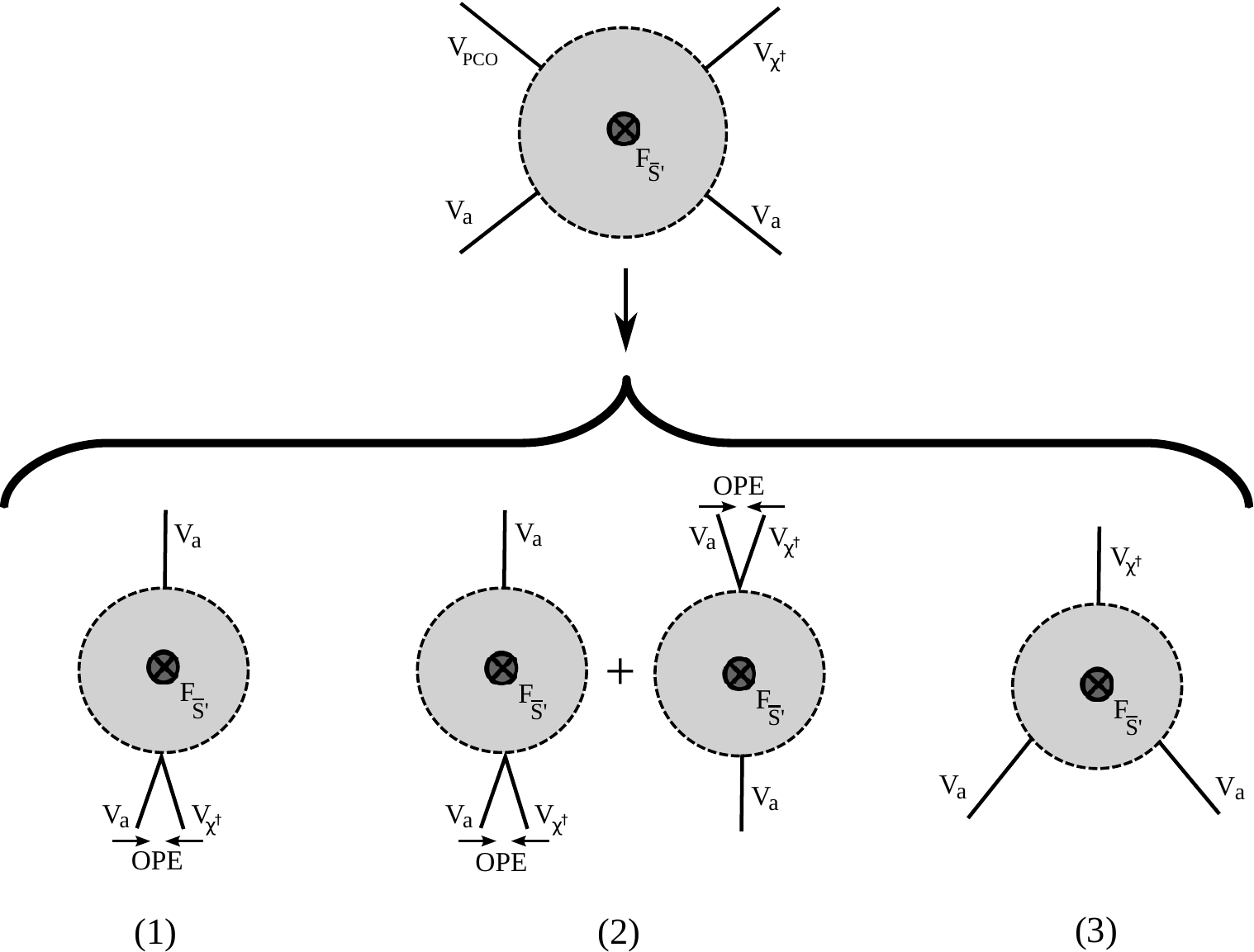,width=13cm}}
\end{center}
\hspace{1cm}\parbox{0.89\textwidth}{\caption{\small{\it Factorisation channels of the disc. The diagram at the top illustrates the four-point function of physical vertices with an insertion of a PCO. The diagrams at the bottom depict various choices for the position of the latter. In case (1), the result takes the form of a contact term, leading to an effective auxiliary field vertex operator insertion on the boundary of the disc. Similarly, case (2) can be re-expressed as a sum of two contact terms, while in case (3) no such interpretation is possible.}}\label{fig-auxiliary}}
\end{figure}

Another way to see that the use of vertex operators ultimately yields the correct physical result is as follows.
Consider the NS-NS contribution \eqref{4ptFctNS} and take the limit where $y$ goes to the points $\bar{z}$, $z$ and $x_2$, respectively, corresponding to using different pictures for the associated operators.  The three cases below are illustrated in Fig. \ref{fig-auxiliary}.

\begin{enumerate}
\item $y\rightarrow \bar{z}$

In this case $A_0\,A_1=0$, whereas $A_0\,A_3$ still cannot produce any pole in the momenta. On the other hand, $A_0\,A_2$ simplifies to
\begin{equation}
\frac{4i p_1\cdot p_3}{x_3-x_1}\frac{z-\bar{z}}{(x_3-z)(x_1-\bar{z})}(\delta^{\nu \lambda}P^{\mu}-\delta^{\mu \lambda}P^{\nu}) \,.
\end{equation}
Notice that the longitudinal mode manifestly decouples. Looking at the vertices (\ref{Va}) and (\ref{Vchi}), we recognize that
$ p_3\cdot p_1/(x_3-x_1)$ appears from contracting $p_1\cdot\chi(x_1)$ with $p_3\cdot\chi(x_3)$. The pole $1/p_1\cdot p_3$ appears from $x_3\rightarrow x_1$. Therefore, in this limit, the result is effectively reproduced by the OPE of the vertices at $x_1$ and $x_3$, resulting in an effective vertex $\Psi\,\psi^\mu$ at $x_1$. This is why in the $(-1,0)$-picture for the NS-NS part of $V_{F^{\bar S'}}$, the auxiliary vertex $V_{Y^\dagger}$ in \eqref{Auxiliary1} gives the correct answer.

\item $y\rightarrow x_2$

In this case again $A_0\,A_1$ vanishes. From $A_0\,A_2$, only the kinematic structure $\delta^{\mu \lambda}P^{\nu}$ survives, with the same answer as above. However the $p_3\cdot p_2/(x_3-x_2)$ factor in $A_0 A_3$ contributes to the other kinematic structure, $\delta^{\nu \lambda}P^{\mu}$. The final result is of course the same but the total result comes from two different factorization limits $x_3 \rightarrow x_1$ and $x_3 \rightarrow x_2$ for the two different kinematic structures.

\item $y\rightarrow z$

In this case $A_0\,A_3$ vanishes, but both the remaining terms contribute. In particular it is not clear if the $A_0\,A_1$ term can even be thought of as a contact term.
\end{enumerate}
Hence, only for $y\rightarrow\bar{z}$ can the result be understood as the factorization in a single channel, such that it can be effectively reproduced by replacing $V_a(x_1)$ and $V_{\chi}(x_3)$
by their OPE, which is simply the auxiliary vertex $V_{Y^\dagger}$ used in Section \ref{Sect:DiskDiagrams}.


Finally, in the R-R contributions \eqref{4ptFctR}, the `contact term' appears only from the channel $x_3 \rightarrow x_1$ and the result can be obtained by a three-point function involving the vertices at $x_2, z, \bar{z}$ and an auxiliary vertex $\Psi\,\psi^\mu$ at $x_1$.  The analysis of the mixed D5-D9 diagrams is very similar and leads to the same conclusion, that is, in the $(-1,0)$-picture for the NS-NS part and $(-1/2,-1/2)$-picture for the RR part of $V_{F^{\bar S'}}$, the entire result comes from contact terms in a single channel and the result can be reproduced by a three-point function involving the auxiliary vertex $V_{\bar{X}^\dagger}$ \eqref{Auxiliary3}.


As mentioned above, even though the calculations are performed on the target space $\mathbb R^4 \times T^2\times\mathbb{R}^4$, the couplings we have obtained are non-vanishing for non-trivial momenta only along the first $\mathbb{R}^4$ (space-time). When we compactify $\mathbb R^4 \times T^2\times\mathbb{R}^4\to \mathbb R^4\times T^2\times K3$, these couplings are unchanged up to possible $\alpha'$ corrections. However, the latter are irrelevant in the field theory limit that we take in order to compare with the non-perturbative part of the $\Omega$-deformed gauge theory partition function. In addition, we have focused here on the gauge theory coming from D9-branes for which the relevant D-instanton is the D5-brane wrapping the internal space. However, it is straightforward to extend our calculation to other setups by applying T-duality. For example, for a gauge theory realised by D5-branes wrapping $T^2$, the relevant D-instanton is the D1-brane wrapped on $T^2$. The corresponding couplings can be obtained from 
the above calculations by performing four T-dualities along the $Y^i$ directions.


\bigskip
\medskip

\bibliographystyle{unsrt}

\vfill\eject

\end{document}